\documentclass[12pt, draftclsnofoot, onecolumn]{IEEEtran}
\usepackage{graphicx}
\usepackage{subfigure}
\usepackage{float}
\usepackage{amsmath}
\usepackage{epstopdf}
\usepackage{cases}
\usepackage{color}
\usepackage{algorithm}
\usepackage{booktabs}
\usepackage{algorithmic}
\usepackage{amsmath}
\usepackage{verbatim}
\usepackage{cite}
\usepackage{lipsum}
\makeatletter
\newenvironment{breakablealgorithm}
{% \begin{breakablealgorithm}
	\begin{center}
		\refstepcounter{algorithm}% New algorithm
		\hrule height.8pt depth0pt \kern2pt% \@fs@pre for \@fs@ruled
		\renewcommand{\caption}[2][\relax]{% Make a new \caption
			{\raggedright\textbf{\ALG@name~\thealgorithm} ##2\par}%
			\ifx\relax##1\relax % #1 is \relax
			\addcontentsline{loa}{algorithm}{\protect\numberline{\thealgorithm}##2}%
			\else % #1 is not \relax
			\addcontentsline{loa}{algorithm}{\protect\numberline{\thealgorithm}##1}%
			\fi
			\kern2pt\hrule\kern2pt
		}
	}{% \end{breakablealgorithm}
	\kern2pt\hrule\relax% \@fs@post for \@fs@ruled
\end{center}
}
\makeatother

\ifCLASSINFOpdf
\else
\fi
\usepackage{caption}
\usepackage{mathtools}

\begin{document}

%
% paper title
% Titles are generally capitalized except for words such as a, an, and, as,
% at, but, by, for, in, nor, of, on, or, the, to and up, which are usually
% not capitalized unless they are the first or last word of the title.
% Linebreaks \\ can be used within to get better formatting as desired.
% Do not put math or special symbols in the title.
%\title{Bare Demo of IEEEtran.cls\\ for IEEE Journals}
\title{Cooperation Based Joint Active and Passive Sensing with Asynchronous Transceivers for Perceptive Mobile Networks}
%
%
% author names and IEEE memberships
% note positions of commas and nonbreaking spaces ( ~ ) LaTeX will not break
% a structure at a ~ so this keeps an author's name from being broken across
% two lines.
% use \thanks{} to gain access to the first footnote area
% a separate \thanks must be used for each paragraph as LaTeX2e's \thanks
% was not built to handle multiple paragraphs
%

%\author{Wangjun~Jiang,~\IEEEmembership{Member,~IEEE,}
%	John~Doe,~\IEEEmembership{Fellow,~OSA,}
%	and~Jane~Doe,~\IEEEmembership{Life~Fellow,~IEEE}% <-this % stops a space
\author{
	Wangjun Jiang,~\IEEEmembership{Student Member,~IEEE,}
	Zhiqing Wei,~\IEEEmembership{Member,~IEEE,} \\
	Shaoshi Yang,~\IEEEmembership{Senior Member,~IEEE,}
	Zhiyong Feng,~\IEEEmembership{Senior Member,~IEEE,} \\
	Ping Zhang,~\IEEEmembership{Fellow,~IEEE}
	\\
	\thanks{This work is supported in part by the National Key Research and Development Program under Grant 2020YFA0711302, in part by the Beijing Municipal Natural Science Foundation under Grant L202012 and Grant Z220004, in part by the Fundamental Research Funds for the Central Universities under Grant 2020RC05, and in part by the BUPT Excellent Ph.D. Students Foundation under Grant CX2022207. \emph{Corresponding author: Zhiyong Feng and Zhiqing Wei.}

	W. Jiang, Z. Wei, S. Yang and Z. Feng are with the School of Information and Communication Engineering, Beijing University of Posts and Telecommunications, and also with the Key Laboratory of Universal Wireless
		Communications, Ministry of Education, Beijing 100876, China (email: \{jiangwangjun, weizhiqing, shaoshi.yang, fengzy\}@bupt.edu.cn).

		P. Zhang is with the School of Information and Communication Engineering, Beijing University of Posts and Telecommunications, and also with the State
		Key Laboratory of Networking and Switching Technology, Beijing 100876,
		China (email: pzhang@bupt.edu.cn).
		}
}

\maketitle

% As a general rule, do not put math, special symbols or citations
% in the abstract or keywords.
\begin{abstract}
%\fontsize{9pt}{6.5pt}
Perceptive mobile network (PMN) is an emerging concept for next-generation wireless networks capable of conducting integrated sensing and communication (ISAC).
A major challenge for realizing high performance sensing in PMNs is how to deal with spatially separated asynchronous transceivers.
Asynchronicity results in timing offsets (TOs) and carrier frequency offsets (CFOs), which further cause ambiguity in ranging and velocity sensing.
%, owing to degraded accuracy in estimating delay and Doppler spread of targets.
Most existing algorithms mitigate TOs and CFOs based on the line-of-sight (LOS) propagation path between sensing transceivers. However, LOS paths may not exist in realistic scenarios. In this paper, we propose a cooperation based joint active and passive sensing scheme for the non-LOS (NLOS) scenarios having asynchronous transceivers. This scheme relies on the cross-correlation cooperative sensing (CCCS) algorithm, which regards active sensing as a reference and mitigates TOs and CFOs by correlating active and passive sensing information.
Another major challenge for realizing high performance sensing in PMNs is how to realize high accuracy angle-of-arrival (AoA) estimation with low complexity. Correspondingly, we propose a low complexity AoA algorithm based on cooperative sensing, which comprises coarse AoA estimation and fine AoA estimation. Analytical and numerical simulation results verify the performance advantages of the proposed CCCS algorithm and the low complexity AoA estimation algorithm.

\end{abstract}

% Note that keywords are not normally used for peerreview papers.

% For peer review papers, you can put extra information on the cover
% page as needed:
% \ifCLASSOPTIONpeerreview
% \begin{center} \bfseries EDICS Category: 3-BBND \end{center}
% \fi
%
% For peerreview papers, this IEEEtran command inserts a page break and
% creates the second title. It will be ignored for other modes.
\IEEEpeerreviewmaketitle

\section{Introduction}
% The very first letter is a 2 line initial drop letter followed
% by the rest of the first word in caps.
%
% form to use if the first word consists of a single letter:
% \IEEEPARstart{A}{demo} file is ....
%
% form to use if you need the single drop letter followed by
% normal text (unknown if ever used by the IEEE):
% \IEEEPARstart{A}{}demo file is ....
%
% Some journals put the first two words in caps:
% \IEEEPARstart{T}{his demo} file is ....
%
% Here we have the typical use of a "T" for an initial drop letter
% and "HIS" in caps to complete the first word.

 \subsection{Background and Motivation}

%Perceptive mobile network (PMN) \cite{[PMC_1],[AOA_9]} is an emerging concept for next-generation wireless networks with the capability of integrated sensing and communication (ISAC).
%PMN is expected to serve as a ubiquitous environment sensing network while providing uncompromising mobile communication services \cite{[PMC]}.
{\color{black}
Perceptive mobile network (PMN) \cite{[PMC_1],[AOA_9]} is an emerging concept for next-generation wireless networks that combine sensing and communication capabilities (ISAC) to serve as a ubiquitous environment sensing network while maintaining reliable mobile communication services \cite{[PMC]}.
}
In \cite{[PMC_3]}, Zhang {\it et al.} conducted an in-depth study on the capability of sensing the user equipment (UE) by the sensing base station (SBS) in PMN, and defined three types of sensing: downlink active sensing, downlink passive sensing, and uplink passive sensing.
In a PMN, active sensing and passive sensing may co-exist \cite{[PMC_3]}. For active sensing, the transmitter and receiver are co-located, which means that they can be conveniently synchronized at the clock level \cite{[jwj_1],[jwj_2]}.
However, for passive sensing, the transmitter and receiver are spatially separated and asynchronous, which results in timing offsets (TOs) and carrier frequency offsets (CFOs), thus leading to degradation of sensing accuracy with respect to ranging and velocity measurements.

To obtain high performance sensing in a PMN, joint active and passive sensing is advocated as a promising technique \cite{[CS_4]}, where cooperation between co-existing active sensing and passive sensing is exploited.
A major challenge facing cooperation based joint active and passive sensing is how to achieve high performance passive sensing with spatially separated asynchronous transceivers.
%理想化场景的研究
On the one hand, most existing passive or cooperative sensing algorithms assumed perfect synchronization between receivers and transmitters, and did not consider the influence of TOs and CFOs on the cooperative sensing performance \cite{[CS_1],[CS_2],[CS_3]}.
%考虑了异步的研究
On the other hand, a few contributions resort to mitigate TOs and CFOs based on the line-of-sight (LOS) path between the sensing-oriented receiver and transmitter, but the LOS path may not exist in realistic scenarios \cite{[Survey],[UL_sensing]}.
%non-LOS场景还是个开放性的问题
For a PMN operating in the non-LOS (NLOS) environment, mitigating the TOs and CFOs in passive sensing remains an open problem that has to be solved.

Since the scale of antenna arrays deployed on SBS is limited, how to achieve high accuracy angle-of-arrival (AoA) estimation with limited-size antenna arrays is another challenge facing the cooperation based joint active and passive sensing.
Most existing algorithms \cite{[UL_sensing],[AOA_29],[AOA_30],[AOA_41]} use both time-domain and spatial-domain measurements to improve the accuracy of AoA estimation, but the computational complexity is excessively high.
Therefore, how to realize high accuracy AoA estimation with low complexity in the context of PMN is also an important problem that needs to be solved.

\subsection{Related Work}
There have been some related studies that are valuable for solving the above mentioned challenges confronting the cooperative sensing in PMN. The major state-of-the-art contributions are described as follows.

1) \textit{TO and CFO mitigation in the presence of asynchronicity:}

In wireless communications, some efforts have been devoted to the asynchronicity problem in cognitive radio \cite{[C_TO_CFO_1],[C_TO_CFO_2],[C_TO_CFO_3]}.
However, these studies focused on the analysis of interference caused by asynchronicity, such as the inter-symbol interference (ISI) and inter-carrier interference (ICI) caused by TOs and CFOs \cite{[C_TO_CFO_4],[C_TO_CFO_5]}, without providing effective methods for parameter estimation.
%Thus, the synchronization accuracy of existing communication synchronization algorithms is difficult to meet the demand of target sensing.

{\color{black}
In wireless sensing, TO and CFO can directly cause timing and Doppler estimation ambiguity and hence leading to degradation of sensing accuracy with respect to ranging and velocity measurements.
Existing synchronization algorithms for sensing can be divided into three categories: GPS clock, and single-node-based and network-based solutions \cite{[AJ_AS]}.
\begin{itemize}
	\item
	GPS clock synchronization is suitable for outdoor environments that can receive GPS signals.
	The synchronization accuracy of GPS-assisted synchronization is sufficient for communications, but not for target sensing.
	%Standard GPS-assisted synchronization is sufficient for communications; however, further processing is required to improve the accuracy and stability of the clock signals for radar sensing applications.
	For example, for a typical GPS clock stability error of 20 ns, this translates into a ranging error of 6 m.
	\item
	Single-node-based synchronization can be implemented in a single receiver. The cross-antenna cross-correlation (CACC) method is a typical single-node-based synchronization algorithm, which is widely used in Wi-Fi sensing \cite{[TO_CFO_25], [TO_CFO_26], [TO_CFO_27]}.
%	Unfortunately, the CACC method has a derivative problem, i.e., its outputs contain multiple mirrored unknown parameters. Recognizing this fact, Ni {\it et al.}\cite{[UL_sensing]} proposed a mirrored multiple signal classification (MUSIC) algorithm to efficiently handle the CACC outputs having equivalently doubled unknown sensing parameters, at the expense of a complexity lower than that of the conventional MUSIC algorithm.
	However, the CACC method mitigate TOs and CFOs based on the assumption that there is a strong LOS path between the sensing receiver and transmitter, which may not be suitable in NLOS scenarios.
	\item Network-based synchronization exploits measurements
	from multiple cooperative nodes. One of the typical network-based synchronization method is the trilateration, which exploit known geometric relationships to remove TO and CFO \cite{[AJ_AS_11]}.
	Other techniques deal with asynchrony by taking advantage of the statistical averaging effect of multiple measurements.
	However, the above methods have the problems of high complexity and difficult multi-target association, which could be concerns for real-time implementation.
\end{itemize}
}
  2) \textit{AoA estimation for wireless sensing:} Another major challenge for realizing high performance sensing in PMNs is how to accurately estimate the AoA of the target.
%  A learning-based AoA estimation method for device-free localization was proposed in \cite{[shen_1]}, which imposes high requirements on the computing power and the scale of antenna arrays.
  AoA estimation algorithms based on the combination of multiple-domain information are proposed in \cite{[AOA_29],[AOA_30],[UL_sensing]}. More specifically, Ni {\it et al.} \cite{[UL_sensing]} attempte to equivalently extend the length of the spatial array response vector by integrating the time and frequency domain signals into the spatial domain, thereby obtaining higher accuracy in AoA estimation.
  In \cite{[AOA_29]}, Chuang {\it et al.} jointly combine spatial and temporal domain information to obtain high-accuracy AoA estimation. In \cite{[AOA_30]}, Ni {\it et al.} define a spatial path filter to separate signals sent over multiple propagation paths and obtain AoA through the CACC output.
  However, the above three algorithms that improve the accuracy of AoA estimation by expanding the length of the spatial array response vector lead to a substantial increase in the computational complexity.

 \subsection{Main Contributions of Our Work}

 Against the above backdrop, we propose a cooperation based joint active and passive sensing scheme for improving the performance of SBS in a PMN that experiences the NLOS propagation, and our focus is on overcoming the asynchronicity problem in passive sensing and the low-complexity high-accuracy AoA estimation challenge with limited scale of antenna arrays. Our major contributions are summarized as follows.

{\color{black}
\begin{enumerate}

\item Considering NLOS propagation, we provide a cooperation based joint active and passive sensing scheme that does not require clock-level synchronization between the spatially separated and asynchronous transceivers. Simulation results show that when the power of the active echo signal is close to that of the passive echo signal, the performance of cooperative sensing becomes appreciably superior to both the active sensing and passive sensing.

\item Considering the single target estimation scenario, we propose a cross-correlation cooperative sensing (CCCS) method that regards active sensing as a reference and mitigates TOs and CFOs existing in passive sensing by correlating active and passive sensing information. Compared with CACC proposed in \cite{[UL_sensing]}, CCCS is more widely applicable, because it does not require the existence of LOS propagation paths between transceivers.

\item Considering the multiple-target estimation scenario, we propose a multi-target alignment algorithm to handle the outputs of the CCCS method, so that the problem of inter-target correlation interference can be solved. Compared with the single target scenario, it is more challenging to estimate TOs and CFOs in the multi-target scenario by directly using CCCS, because in this situation the outputs of CCCS not only contain information on TOs and CFOs of the same target, but also contain information on the delay and Doppler spread correlation between different targets. This correlation may hamper the subsequent TOs and CFOs mitigation operations. The multi-target alignment algorithm includes two stages, i.e., the extraction stage and the matching stage. In the extraction stage, TOs and CFOs of the same target can be extracted with the aid of spatial information differences of different targets. Then TOs and CFOs of different targets are matched in the matching stage, as detailed in Section \ref{sec:Delay_Doppler_4_2}. Simulation results demonstrate that the multi-target alignment algorithm works well when the variance of CFOs is smaller than $\frac{1}{4}$ of the subcarrier spacing, which can be satisfied by ordinary frequency sources \cite{[CFO_S_3],[CFO_S_4],[CFO_S_5]}.

\item We develop a low-complexity high-accuracy AoA estimation algorithm based on cooperative sensing and fractional Fourier transform (FRFT). Although traditional FRFT can significantly improve the accuracy of AoA estimation, its way of expanding the number of Fourier transform points increases its computational complexity \cite{[QHY]}.
To reduce the complexity while ensuring high-accuracy AoA estimation, we propose an AoA estimation algorithm relying on iterations between coarse estimation and fine estimation.
In the coarse estimation stage, we obtain a rough AoA result of targets by cooperative active and passive sensing. In the fine estimation stage, we further obtain a fine AoA of targets by the FRFT algorithm within the range of the rough AoA result.
We analyze and simulate the accuracy and complexity of different AoA estimation algorithms, including the hybrid multiple signal classification (H-MUSIC) \cite{[AOA_29]}, hybrid estimation of signal parameters via rotational invariance techniques (H-ESPRIT) \cite{[AOA_29]}, joint time-space-frequency (TSF) domain MUSIC algorithm (TSF-MUSIC) \cite{[UL_sensing]}, and FRFT \cite{[QHY]}.
simulation results show that the AoA algorithm proposed in this paper can achieve relatively high-performance AoA estimation, close to FRFT, at a low complexity, much lower than FRFT.
%
%
%To reduce the complexity while ensuring high-accuracy AoA estimation, we propose an AoA estimation algorithm relying on iterations between coarse estimation and fine estimation.
%In the coarse estimation stage, we obtain a coarse AoA of targets by the DFT algorithm. Based on the coarse AoA, we further obtain a fine AoA of targets by the TSF-MUSIC algorithm, which integrates both the time domain and the frequency domain signals into the spatial domain.
%
%Simulation results show that our AoA estimation algorithm is capable of achieving high-accuracy AoA estimation, close to the TSF-MUSIC in \cite{[UL_sensing]}, at a low complexity, close to the hybrid MUSIC (H-MUSIC) in \cite{[AOA_29]}.
\end{enumerate}
}

The rest of this paper is organized as follows.
In Section \ref{sec:system_model} we describe the system model of the cooperation based joint active and passive sensing. The cooperative sensing for delay and Doppler spread is introduced in Section \ref{sec:Delay_Doppler}. In Section \ref{sec:AOA} we present the proposed low-complexity high-accuracy AoA estimation algorithm. Range, velocity and AoA estimation are analyzed and numerically evaluated in Section \ref{sec:Simulation}. Finally, Section \ref{sec:Conclusion} concludes the paper.

\textit{Notations}: Vectors and matrices are denoted by boldface lowercase and uppercase letters; the transpose, complex conjugate, Hermitian, inverse, and pseudo-inverse of the matrix ${\bf{A}}$ are denoted by ${{\bf{A}}^{\rm T}}$, ${{\bf{A}}^*}$, ${{\bf{A}}^{\rm H}}$, ${{\bf{A}}^{ - 1}}$ and ${{\bf{A}}^\dag}$, respectively; ${\rm{diag}}(\bf{x})$ denotes a diagonal matrix whose diagonal elements are the elements of $\bf x$; ${\mathcal S}(x_i)$ denotes the set of $x_i$; $ \otimes $ represents the Kronecker product; $ \odot $ denotes the Hadamard product.

\section{System Model of Cooperative Sensing}\label{sec:system_model}

%\begin{figure}[ht]
%	\includegraphics[scale=0.35]{MSBS/system_model_4.eps}
%	\centering
%	\caption{System model of cooperative sensing.}
%	\label{fig:ISAC_system_model}
%\end{figure}

\begin{figure*}[ht]
	\centering
	\subfigure[\scriptsize{SBS senses vehicle targets.}]{\includegraphics[width=0.42\textwidth]{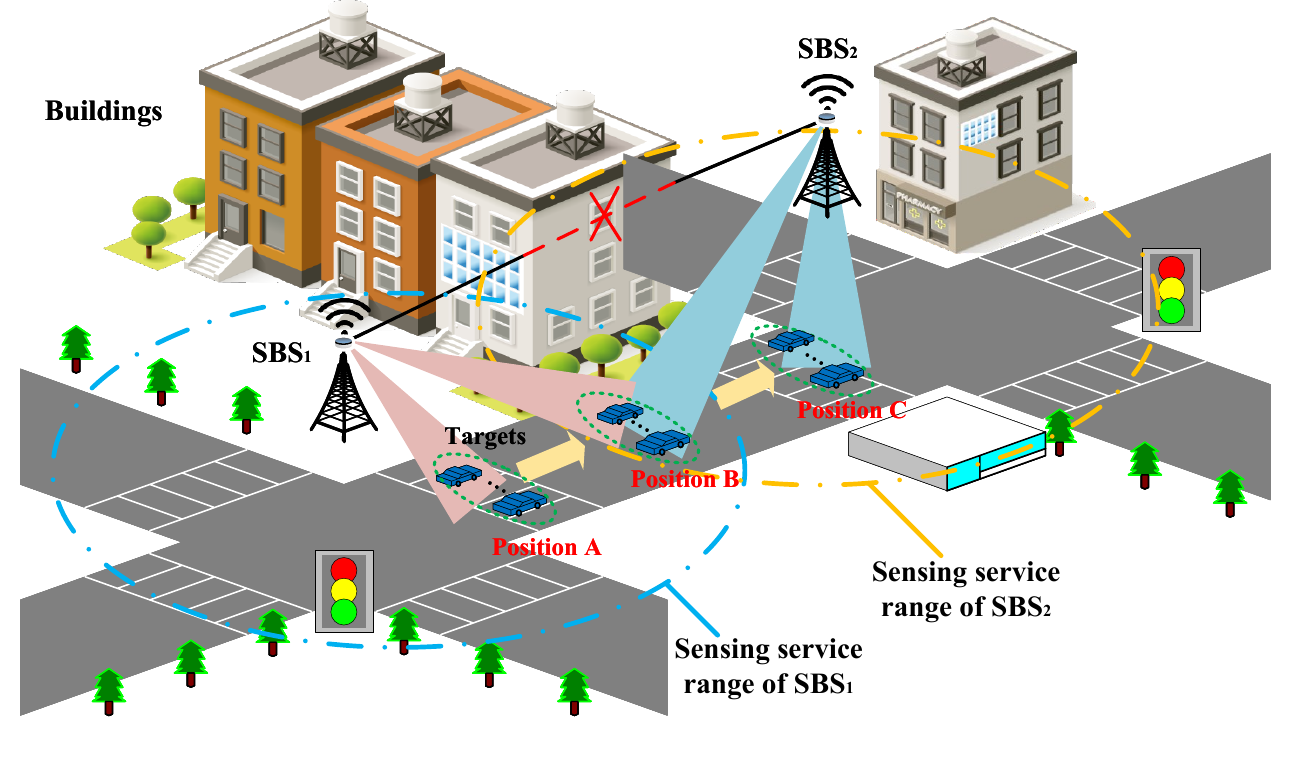}%
		\label{fig:system_1}}
	\hfil
	\subfigure[\scriptsize{Active and passive sensing}]{\includegraphics[width=0.49\textwidth]{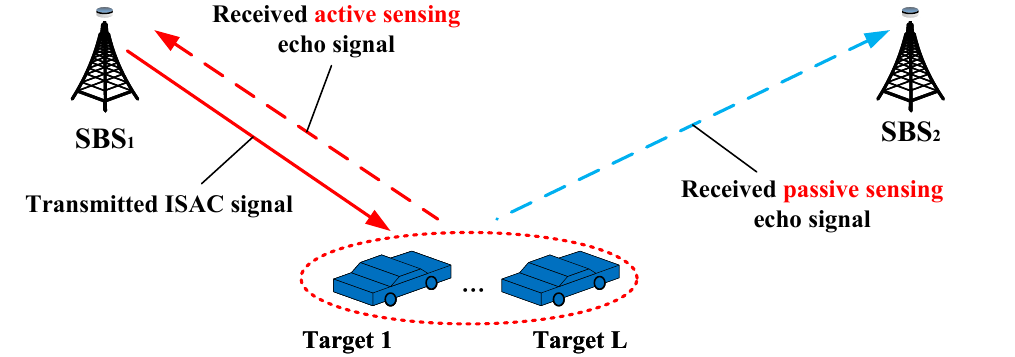}%
		\label{fig:system_2}}
	\hfil
	\caption{System model of cooperative sensing.}
	\label{fig:ISAC_system_model}
\end{figure*}

We consider the cooperation based joint active and passive sensing in a PMN. As Fig.~ \ref{fig:ISAC_system_model} shows there is no LOS propagation path between $\rm{SBS}_1$ and $\rm{SBS}_2$ due to the occlusion of buildings.
Each SBS with $N_t$ antennas, transmits ISAC signal to targets. These targets are not only sensing targets but also communication users.
On the one hand, these targets obtain communication services by demodulating ISAC signals sent from SBS.
On the other hand, SBS can receive the echo signal reflected by targets to realize the target sensing.
These targets move along the road, across the sensing service range of $\rm{SBS}_1$ and $\rm{SBS}_2$.
To provide continuous and high-performance sensing services for these targets, we propose a collaboration scheme between $\rm{SBS}_1$ and $\rm{SBS}_2$.
% , including active perception and active passive collaborative perception
{\color{black}
	It should be noted that, in this paper, we focuses on the use of ISAC technology to empower conventional communication base stations with sensing capabilities.
	\\
	\begin{itemize}
		\item In terms of signal design, we just reuses the OFDM communication signals of the existing base stations to achieve sensing without modification, and therefore, there is no significant interference to the communication performance.
		\item In terms of communication interaction, we just uses the echo signal reflected from the base station after the target for active and passive sensing, without destroying the original communication link, therefore, there is no obvious interference to the communication performance.
        Compared to the traditional communication base station, one of the biggest differences is that the ISAC base station needs to deploy two antenna arrays to simultaneously complete the ISAC signal transmission and echo signal reception. This can be done by increasing the cost of hardware to mitigate the degradation of communication performance.
	\end{itemize}
    Therefore, we think that the reduction in communication performance can be compensated by the use of appropriate technology and hardware cost inputs, and in this paper we do not focus on analyzing the specific performance of the communication link.
    There have been related studies on how to design ISAC waveforms to trade-off sensing and communication performance \cite{[CD_OFDM]}, which is not the focus of this paper and may be further investigated in future work.
}

\subsection{Cooperation between $\rm{SBS}_1$ and $\rm{SBS}_2$}\label{sec:system_model_1}

{\color{black}
As Fig. \ref{fig:system_1} shows, when targets in position A, in the sensing service range of $\rm{SBS}_1$, $\rm{SBS}_1$ can provide active sensing services for targets; when targets in position C, in the sensing service range of $\rm{SBS}_2$, $\rm{SBS}_2$ can provide active sensing services for targets. However, when targets in position B, in the sensing service overlap range of $\rm{SBS}_1$ and $\rm{SBS}_2$, the power of active sensing echo signal is weak, making it difficult for $\rm{SBS}_1$ and $\rm{SBS}_2$ to provide high-performance active sensing services for targets.
We first use passive sensing to make up for the lack of performance of active sensing. However, the transceiver nodes of passive sensing are spatially separated asynchronously, resulting in TOs and CFOs, which further cause ambiguity in ranging and velocity sensing.
To solve the above problem, we provide a cooperation based joint active and passive sensing scheme, including the following steps:
\begin{itemize}
	\item Parameter setting:
	$\rm{SBS}_1$ and $\rm{SBS}_2$ adopt different signal parameters to distinguish the received active and passive sensing echo signals, as shown in Fig. \ref{fig:system_2}.
	The techniques for separating echo signals of adjacent base stations are relatively mature, including time division (TD) scheme, frequency division (FD) scheme, code division (CD) scheme, etc.
	Although it is easier to distinguish between active and passive sensing signals in the TD scheme, active and passive sensing of the target cannot be simultaneously achieved by the SBS in this case \cite{[TDD_1],[TDD_2],[TDD_4]}.
	By contrast, in the CD scheme, SBS can achieve active and passive sensing of the target at the same time, but it requires complex coding design and signal processing to distinguish between active and passive sensing signals \cite{[TDD_3],[CD_OFDM]}.
	In this paper, under the condition of sufficient spectrum resources, we assume that adjacent SBSs adopt the FD scheme \cite{[FDD_31]}, which is widely adopted for adjacent base stations in the $4$-th generation (4G) and the $5$-th generation (5G) mobile communication systems \cite{[3GPP_FD]}.
	\item Initiation of cooperative sensing:
	When $\rm{SBS}_1$ receives the echo signal sent from 	$\rm{SBS}_2$, $\rm{SBS}_1$ will send a cooperative sensing request to $\rm{SBS}_2$. After receiving the reply of $\rm{SBS}_2$, the two SBSs will start cooperative sensing at the same time to generate a sensing beam pointing to the target area.
	\item Signal processing of cooperative sensing:
	When $\rm{SBS}_1$ receives the active and passive sensing echo signals, it will perform cooperative sensing based on CCCS, which will be described in detail in Section \ref{sec:Delay_Doppler}.
\end{itemize}

}

\subsection{Received Signal Model of SBS}\label{sec:system_model_2}
{\color{black}
	SBSs in a PMN can provide both sensing and communication services to UEs by transmitting ISAC signals.
    {\color{black}
    In the ISAC system, ISAC base stations not only need to improve sensing services, but also need to provide communication services for communication users.
    If the conventional sensing waveforms are used, it is difficult to guarantee the performance of the communication link. Therefore, the orthogonal frequency division multiplexing (OFDM) signal was used for joint sensing and communication \cite{[OFDM_12]}.
}
	In terms of communication, OFDM signal is the waveform of the 4G and 5G mobile communication systems, hence it not only has advantages in mitigating multipath interference and exploiting frequency diversity, but also enjoys standard compatibility \cite{[OFDM_2]}.
	In terms of sensing, OFDM signal has advantages in accuracy, resolution and flexibility \cite{[OFDM_advantage]}. The number and the spacing of subcarriers can be adjusted flexibly to obtain the ambiguity function of ``pushpin shape" \cite{[OFDM_3]}.
	Therefore, OFDM signal is adopted as the sensing waveform of the SBS in this paper.
	%OFDM signal is a typical signal model of base station sensing \cite{[OFDM],[PMC]}, which is used for SBS sensing in this paper.
	Then, the received signal of ${\rm SBS}_1$ can be expressed as
	\begin{equation}\label{equ:Received_Signal_Model_1}
		\begin{aligned}
			{ y}(t)  &= \sum_{k=0}^{N_t - 1}   {y}_1(t) + {y}_2(t) + {z}(t) ,  \\
			%		{\bf a}({\Omega}_{l}) &= [1, e^{j{\Omega}_{l}},\cdots,e^{j{\Omega}_{l}k},\cdots,e^{j(N_t-1){\Omega}_{l}}]^\textrm{T} \in {{\mathcal C}^{N_t \times 1}}
		\end{aligned}
	\end{equation}
	where ${z}(t)$ is a complex additive white Gaussian noise (AWGN) vector with zero mean and variance of $\sigma^2$, ${y}_1(t) $ and ${y}_2(t)$ denote the echo signals from ${\rm SBS}_1$ and ${\rm SBS}_2$, respectively. The notation is defined as follows:
	\begin{itemize}
		\item $k$ is the index of antenna elements.
	\end{itemize}
	The active sensing echo signal from ${\rm SBS}_1$ and the passive sensing echo signal from ${\rm SBS}_2$ can be expressed as \cite{[MIMO-OFDM],[MIMO-OFDM_GFF_1],[MIMO-OFDM_GFF_2],[MIMO-OFDM_JS_1],[MIMO-OFDM_JS_2],[MIMO-OFDM_WXD_1]}
	\begin{equation} \label{equ:Received_Signal_Model_2}
		\begin{aligned}
			{{y}_1(t)}  &= \sum_{m=0}^{M-1} \sum_{n=0}^{N-1} \overbrace { d_{1,\textrm{Tx}}(k,m,n) \times \sum_{l=0}^{L-1} \alpha_{1,l} e^{j{\Omega}_{l}k}   e^{-j2 \pi n \Delta f {\tau_{1,l}}} \times e^{j2 \pi m T f_{D,1,l}}  }^{d_{1,\textrm{Rx}}(k,m,n)}
		\end{aligned}
	\end{equation}
	\begin{equation} \label{equ:Received_Signal_Model_3}
		\begin{aligned}
			& \cdot e^{-j 2 \pi (f_\textrm{c1} + n \Delta f) t} \cdot {\rm rect} \left(\frac{t-mT-\tau_{1,l}} {T}\right),
		\end{aligned}
	\end{equation}

	\begin{equation} \label{equ:Received_Signal_Model_4}
		\begin{aligned}
			{{y}_2(t)}  &= \sum_{m=0}^{M-1} \sum_{n=0}^{N-1} \overbrace { d_{2,\textrm{Tx}}(k,m,n) \times \sum_{l=0}^{L-1} \alpha_{2,l} e^{j{\Omega}_{l}k} e^{-j2 \pi n \Delta f (\tau_{2,l}+\delta_\tau(m))} \times e^{j2 \pi m T (f_{D,2,l} + \delta_f(m))} }^{d_{2,\textrm{Rx}}(k,m,n)}
		\end{aligned}
	\end{equation}
	\begin{equation} \label{equ:Received_Signal_Model_5}
		\begin{aligned}
			\cdot e^{-j 2 \pi (f_\textrm{c2} + n \Delta f) t} \cdot {\rm rect} \left(\frac{t-mT-\tau_{2,l}-{\delta_\tau(m)}} {T}\right),
		\end{aligned}
	\end{equation}
	where ${\rm rect} \{\cdot\}$ denotes a rectangular window function. The rest of the notations and terms can be defined as follows:
	\begin{itemize}
		\item ${\bf a}({\Omega}_{l})$ is the receive steering vector of the $l$-th target with ${\Omega}_{l} = \frac{2 \pi d}{\lambda } {\rm{cos}}(\theta_l )$,
		\item $d$ is the distance between adjacent antenna elements,
		\item ${\lambda }$ is the wavelength of the signal,
		\item $\theta_l$ is the direction of the $l$-th target,
		\item $M$ and $N$ represent the number of OFDM symbols and the number of subcarriers,
		\item $m$ and $n$ are the indices of the OFDM symbols and the subcarriers, respectively,
		\item $f_\textrm{c1}$ and $f_\textrm{c2}$ are the carrier frequency of the active sensing and passive sensing echo signal,
		\item $L$ is the number of targets,
		\item $T$ is the length of an OFDM symbol,
		\item $\Delta f$ is the carrier spacing of OFDM signals.
		\item $d_{1,\textrm{Rx}}(k,m,n)$ and $d_{2,\textrm{Rx}}(k,m,n)$ are the received modulation symbols of the $k$-th antenna on $\rm{SBS}_1$ and $\rm{SBS}_2$,
		\item $d_{1,\textrm{Tx}}(k,m,n)$ and $d_{2,\textrm{Tx}}(k,m,n)$ are the transmitted modulation symbols of the $k$-th antenna on $\rm{SBS}_1$ and $\rm{SBS}_2$,
		\item $\tau_{1,l}$ and $f_{D,1,l}$ are the delay and Doppler spread between the $l$-th target and $\rm{SBS}_1$,
		\item $\tau_{2,l}$ and $f_{D,2,l}$ are the delay and Doppler spread between the $l$-th target and $\rm{SBS}_2$,
		\item $\alpha_{1,l}$ and $\alpha_{2,l}$ are the channel fading magnitudes of the $l$-th target in active and passive sensing,
		\item $\delta_\tau(m)$ and $\delta_f(m)$ denote the unknown time-varying TO and CFO \cite{[TO_CFO_1],[TO_CFO_26]}.
	\end{itemize}
	%Furthermore, the $m$-th OFDM symbol and the $n$-th subcarrier received from $\rm{SBS}_1$ and $\rm{SBS}_2$ can be expressed as
	%%\begin{equation} \label{equ:Received_Signal_Model_4}
	%%\begin{aligned}
	%%d_{1,\textrm{Rx}}(k,m,n) &= d_{1,\textrm{Tx}}(k,m,n) \times \sum_{l=0}^{L-1} \alpha_{1,l} e^{j{\Omega}_{l}k}   e^{-j2 \pi n \Delta f {\tau_{1,l}}} \times e^{j2 \pi m T f_{D,1,l}},
	%%\end{aligned}
	%%\end{equation}
	%%\begin{equation} \label{equ:Received_Signal_Model_5}
	%%\begin{aligned}
	%%d_{2,\textrm{Rx}}(k,m,n) &= d_{2,\textrm{Tx}}(k,m,n) \times \sum_{l=0}^{L-1} \alpha_{2,l} e^{j{\Omega}_{l}k} e^{-j2 \pi n \Delta f (\tau_{2,l}+\delta_\tau(m))} \times e^{j2 \pi m T (f_{D,2,l} + \delta_f(m))},
	%%\end{aligned}
	%%\end{equation}
	%where
	%\begin{itemize}
	%	\item $d_{1,\textrm{Tx}}(k,m,n)$ and $d_{2,\textrm{Tx}}(k,m,n)$ are the transmitted modulation symbols of the $k$-th antenna on $\rm{SBS}_1$ and $\rm{SBS}_2$,
	%	\item $\tau_{1,l}$ and $f_{D,1,l}$ are the delay and Doppler spread between the $l$-th target and $\rm{SBS}_1$,
	%	\item $\tau_{2,l}$ and $f_{D,2,l}$ are the delay and Doppler spread between the $l$-th target and $\rm{SBS}_2$,
	%	\item $\alpha_{1,l}$ and $\alpha_{2,l}$ are the channel fading magnitudes of the $l$-th target in active and passive sensing,
	%	\item $\delta_\tau(m)$ and $\delta_f(m)$ denote the unknown time-varying TO and CFO \cite{[TO_CFO_1],[TO_CFO_26]}.
	%\end{itemize}
}

\section{Delay and Doppler Spread Estimation Based on CCCS}\label{sec:Delay_Doppler}

%描述TO和CFO问题
According to Section \ref{sec:system_model_2}, the delay and Doppler spread are mixed with TOs and CFOs, respectively. In this section, we propose the CCCS based algorithm to mitigate TOs and CFOs for achieving {\color{black} high-accuracy} target delay and Doppler spread estimation.
The CCCS based algorithm for delay and Doppler spread estimation consists of the following steps.

%\subsection{Matrix Representation of Modulation Symbols}\label{sec:Delay_Doppler_1}
%For convenience of representation, the modulation symbol frame of the transmitted and received signals is rewritten as a matrix, in which each column represents a OFDM symbol and each row represents a subcarrier. Then, the modulation symbol matrix ${\bf D}_{1,\textrm{Tx}}(k)$ transmitted from the $k$th antenna of $\rm{SBS}_1$ can be expressed as
%\begin{equation} \label{equ:CSCC_1}
%\begin{aligned}
%&{\bf D}_{1,\textrm{Tx}}(k) = \begin{bmatrix}
%d_{1,\textrm{Tx}}(k,0,0)&\cdots &d_{1,\textrm{Tx}}(k,M-1,0) \\
%\vdots & \ddots  &\vdots \\
%d_{1,\textrm{Tx}}(k,0,N-1)&\cdots  &d_{1,\textrm{Tx}}(k,M-1,N-1)
%\end{bmatrix}
%%, \in {{\mathcal C}^{N \times M}}
%\end{aligned}.
%\end{equation}
%The transmitted modulation symbol matrix ${\bf D}_{2,\textrm{Tx}}$ from $\rm{SBS}_2$ and received modulation symbol matrix ${\bf D}_{1,\textrm{Rx}}$, ${\bf D}_{2,\textrm{Rx}}$ from $\rm{SBS}_1$ and $\rm{SBS}_2$ can be regarded in the same matrix representation.

\subsection{Extracting Sensing Information}\label{sec:Delay_Doppler_2}

{\color{black} Based on \eqref{equ:Received_Signal_Model_1}, \eqref{equ:Received_Signal_Model_4} and \eqref{equ:Received_Signal_Model_5}, the transmitted and received modulation symbols of the $k$-th antenna array from $\rm{SBS}_1$ and $\rm{SBS}_2$ can be rewritten as matrices, ${\bf D}_{1,\textrm{Tx}}(k)$, ${\bf D}_{2,\textrm{Tx}}(k)$, ${\bf D}_{1,\textrm{Rx}}(k)$, ${\bf D}_{2,\textrm{Rx}}(k)$, in which each column represents a OFDM symbol and each row represents a subcarrier. }
The sensing information can be extracted from the received modulation symbols by a point-wise complex division. {\color{black} Since the transmitted signal is modulated by quadrature phase shift keying (QPSK) in this paper, the amplitude of the transmitted OFDM modulation symbol is a non-zero constant, the noise enhancement caused by point-wise complex division is not serious \cite{[CD_OFDM]}. Without loss of generality, we take the example of extracting sensing information from ${\bf D}_{1,\textrm{Rx}}(k)$ \cite{[OFDM],[WK-OFDM]}
}
\begin{equation} \label{equ:CSCC_2}
\begin{aligned}
{\bf D}_\textrm{div}(k) &= \frac{{\bf D}_{1,\textrm{Rx}}(k)}{{\bf D}_{1,\textrm{Tx}}(k)} =
% \approx
{\bf k}_{1,{R}}(k) \otimes {\bf k}_{1,{D}}(k)
%, \in {{\mathcal C}^{N \times M}}
\end{aligned},
\end{equation}
where $ \otimes $ is the Kronecker product operator and
\begin{equation} \label{equ:CSCC_3}
\begin{aligned}
{\bf k}_{1,{R}}(k) &= \sum_{l=0}^{L-1} \alpha_{1,l} e^{j{\Omega}_{l}k} { \begin{bmatrix}
1, & \cdots, & e^{-j2 \pi n \Delta f \tau_{1,l}}, & \cdots, &   e^{-j2 \pi (N-1) \Delta f \tau_{1,l}}
	\end{bmatrix} }^\textrm{T}
%, \in {{\mathcal C}^{N \times 1}}
\end{aligned},
\end{equation}
\begin{equation} \label{equ:CSCC_4}
\begin{aligned}
{\bf k}_{1,{D}}(k) &= \sum_{l=0}^{L-1} \alpha_{1,l} e^{j{\Omega}_{l}k}
{ \begin{bmatrix} 1,& \cdots,& e^{j2 \pi m T f_{D,1,l}},& \cdots,&   e^{j2 \pi (M-1) T f_{D,1,l}}  \end{bmatrix} }^\textrm{T}
%, \in {{\mathcal C}^{M \times 1}}
\end{aligned},
\end{equation}
are the range and Doppler spread steer vectors of active sensing.

Similarly, ${\bf k}_{2,{R}}(k)$ and ${\bf k}_{2,{D}}(k)$ can be obtained by extracting sensing information from from ${\bf D}_{2,\textrm{Rx}}$
\begin{equation} \label{equ:CSCC_5}
	\begin{aligned}
		{\bf k}_{2,{R}}(k) &= \sum_{l=0}^{L-1} \alpha_{2,l} e^{j{\Omega}_{l}k}
		{\begin{bmatrix}  1,& \cdots,& e^{-j2 \pi n \Delta f (\tau_{2,l} + \delta_\tau(m))},& \cdots,&   e^{-j2 \pi (N-1) \Delta f (\tau_{2,l} + \delta_\tau(m))}  \end{bmatrix} }^\textrm{T}
	\end{aligned},
\end{equation}
\begin{equation} \label{equ:CSCC_6}
\begin{aligned}
{\bf k}_{2,{D}}(k) &= \sum_{l=0}^{L-1} \alpha_{2,l} e^{j{\Omega}_{l}k}
{\begin{bmatrix} 1,& \cdots,& e^{j2 \pi m T (f_{D,2,l} + \delta_f(m))},&  \cdots,&   e^{j2 \pi (M-1) T (f_{D,2,l} + \delta_f(m))}
	\end{bmatrix} }^\textrm{T}
\end{aligned}.
\end{equation}
\subsection{Deviation Analysis of Cooperative Active and Passive Sensing}\label{sec:Delay_Doppler_3}

\begin{figure}[ht]
	\includegraphics[scale=0.65]{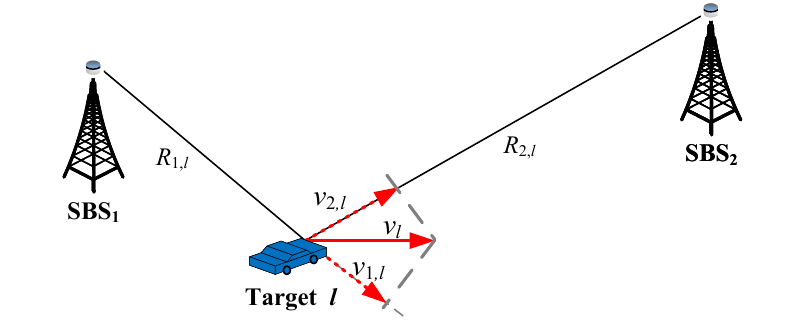}
	\centering
	\caption{Active and passive sensing for the $l$-th target.}
	\label{fig:DA}
\end{figure}
Due to the different location information of $\rm{SBS}_1$ and $\rm{SBS}_2$, there are deviations in the delay and Doppler spread estimation of active and passive sensing for the same target, which needs to be mitigated in order to achieve high accurate cooperative sensing.
Without loss of generality, {\color{black} we consider the active and passive sensing for the $l$-th target.} As Fig. \ref{fig:DA} shows, the distance from the $l$-th target to $\rm{SBS}_1$ and $\rm{SBS}_2$ is $R_{1,l}$ and $R_{2,l}$, the velocity of target in the direction of $\rm{SBS}_1$ and $\rm{SBS}_2$ is $v_{1,l}$ and $v_{2,l}$. {\color{black} The delay, $\tau_{1,l}$ and $\tau_{2,l}$ can be expressed as}
\begin{equation} \label{equ:CSCC_7}
\begin{aligned}
\tau_{1,l} = \frac{2R_{1,l}}{c}, \quad
\tau_{2,l} = \frac{R_{1,l} + R_{2,l}}{c}
\end{aligned},
\end{equation}
where $c$ is the speed of signal, $R_{1,l}$ and $R_{2,l}$ are the range from the $l$-th target to $\rm{SBS}_1$ and $\rm{SBS}_2$, respectively.
Then the deviation of active and passive sensing in terms of delay estimation is
\begin{equation} \label{equ:CSCC_8}
\begin{aligned}
\Delta \tau_{l} &= \tau_{2,l} - \tau_{1,l} = \frac{R_{2,l} - R_{1,l}}{c}
\end{aligned}.
\end{equation}
{\color{black}
Similarly, the Doppler spread, $f_{{D},1,l}$ and $f_{{D},1,l}$ can be expressed as
\begin{equation} \label{equ:CSCC_9}
	\begin{aligned}
		f_{{D},1,l} = \frac{2v_{1,l} f_\textrm{c1}}{c}, \quad
		f_{{D},2,l} = \frac{(v_{1,l} - v_{2,l}) f_\textrm{c2}}{c}
	\end{aligned},
\end{equation}
and the deviation of active and passive sensing in terms of Doppler spread estimation $\Delta f_{{D},l}$ can be derived as
\begin{equation} \label{equ:CSCC_10}
	\begin{aligned}
		\Delta f_{{D},l} &= f_{{D},2,l} - f_{{D},1,l} = \frac{v_{1,l} (f_\textrm{c2} - f_\textrm{c1}) }{c} - \frac{(v_{1,l} + v_{2,l}) f_\textrm{c2}}{c}
	\end{aligned},
\end{equation}
where $v_{1,l}$ and $v_{2,l}$ are the velocity of the $l$-th target in the direction from $\rm{SBS}_1$ to the $l$-th target and from the $l$-th target to $\rm{SBS}_2$.
}

\subsection{CCCS for Mitigating TOs and CFOs}\label{sec:Delay_Doppler_4}
To obtain better sensing performance by fusing the sensing signals of active and passive sensing, it is necessary to mitigate TOs and delay deviation of passive sensing in range measurement, and to mitigate the CFOs and Doppler spread deviation of passive sensing in velocity measurement.
The CCCS based algorithm is proposed to mitigate TOs plus delay deviation and CFOs plus Doppler spread deviation by correlating active and passive sensing information.
{\color {black}
Specifically, the CCCS based algorithm regards the active sensing echo signal as a reference signal, and mitigates TOs and CFOs in the passive sensing echo signal by correlating the active and passive echo signals, which is named the CCCS algorithm.
}

\subsubsection{Single target}\label{sec:Delay_Doppler_4_1}
For single target, i.e., $L = 1$, we perform the same CCCS algorithm for the received signal on each antenna. Without loss of generality, taking the $0$-th antenna as an example, i.e., $k = 0$, then $e^{j \Omega_l k} = 1$.
%since the channel fading amplitudes $\alpha_{1}$ and $\alpha_{2}$ do not affect the mitigation of TOs and CFOs,
Then, the active and passive sensing information normalized vectors can be re-expressed as
\begin{equation} \label{equ:CSCC_4_1}
\begin{aligned}
{\bf k}_{1,{R}} &=
{\begin{bmatrix}  1,& \cdots,&   e^{-j2 \pi (N-1) \Delta f \tau_{1}}  \end{bmatrix} }^\textrm{T}, \qquad \qquad
{\bf k}_{1,{D}} =
{\begin{bmatrix}  1,& \cdots &   e^{j2 \pi (M-1) T f_{{D},1}}  \end{bmatrix} }^\textrm{T}, \\
{\bf k}_{2,{R}} &=
{\begin{bmatrix}  1,& \cdots,&   e^{-j2 \pi (N-1) \Delta f (\tau_{2} + \delta_\tau(m))}  \end{bmatrix} }^\textrm{T},  \quad
{\bf k}_{2,{D}} =
{\begin{bmatrix}  1,& \cdots,&   e^{j2 \pi (M-1) T (f_{{D},2} + \delta_f(m))}  \end{bmatrix} }^\textrm{T}
\end{aligned}.
\end{equation}
Deviations of delay and Doppler spread between active and passive sensing for the same target can be re-expressed as $\Delta \tau$ and $\Delta f_{{D}}$, respectively.
For mitigating $\Delta \tau + \delta_\tau(m)$, the CCCS algorithm between ${\bf k}_{1,{R}}$ and ${\bf k}_{2,{R}}$ generates
\begin{equation} \label{equ:CSCC_4_2}
\begin{aligned}
{\bf \rho}_{{R}} &=  {\rm{diag}} ({\bf k}_{2,{R}}^{\textrm H}) \cdot {\bf k}_{1,{R}}
&=
{\begin{bmatrix}  1,& \cdots,& e^{j2 \pi n \Delta f (\Delta \tau + \delta_\tau(m))}, & \cdots,&   e^{j2 \pi (N-1) \Delta f (\Delta \tau + \delta_\tau(m))} \end{bmatrix} }^\textrm{T}
%, \in {{\mathcal C}^{N \times 1}}
\end{aligned}.
\end{equation}
For mitigating $\Delta f_{{D}} + \delta_f(m)$, the CCCS algorithm between ${\bf k}_{1,{D}}$ and ${\bf k}_{2,{D}}$ generates
\begin{equation} \label{equ:CSCC_4_3}
\begin{aligned}
{\bf \rho}_{{D}} &=   {\rm{diag}} ({\bf k}_{2,{D}}^{\textrm H})\cdot  {\bf k}_{1,{D}}
&=
{\begin{bmatrix}  1,& \cdots,& e^{-j2 \pi m T (\Delta f_{{D}} + \delta_f(m))},& \cdots,&   e^{-j2 \pi (M-1) T (\Delta f_{{D}} + \delta_f(m))} \end{bmatrix} }^\textrm{T}
%, \in {{\mathcal C}^{M \times 1}}
\end{aligned}.
\end{equation}
where $\Delta \tau + \delta_\tau(m)$ translates into a linear phase shift between the modulation symbols along the carrier frequency axis, $\Delta f_{{D}} + \delta_f(m)$ translates into a linear phase shift between the modulation symbols along the OFDM symbol axis.
Thus, $\Delta \tau + \delta_\tau(m)$ and $\Delta f_{{D}} + \delta_f(m)$ can be estimated by the discrete Fourier transform (DFT) algorithm \cite{[OFDM]}.

%jwj
%Then, the passive sensing information vector after mitigating $\Delta \tau + \delta_\tau(m)$ can be derived as
%\begin{equation} \label{equ:CSCC_4_4}
%\begin{aligned}
%{\bf k}_{2,\textrm{R}}^{'} &=  {\rm{diag}} ({\bf \rho}_{\textrm{R}}^{\textrm H}) \cdot {\bf k}_{2,\textrm{R}}
%&=  {\begin{bmatrix}  1,& \cdots,& e^{-j2 \pi n \Delta f \tau_{1}},& \cdots,&   e^{-j2 \pi (N-1) \Delta f \tau_{1}}  \end{bmatrix} }^\textrm{T}
%& = {\bf k}_{1,\textrm{R}}
%\end{aligned}.
%\end{equation}
%The passive sensing information vector after mitigating $\Delta \tau + \delta_\tau(m)$ can be derived as
%\begin{equation} \label{equ:CSCC_4_5}
%\begin{aligned}
%{\bf k}_{2,\textrm{D}}^{'} &= {\rm{diag}} ({\bf \rho}_{\textrm{D}}^{\textrm H}) \cdot {\bf k}_{2,\textrm{D}}
%&=  {\begin{bmatrix}  1,& \cdots,& e^{j2 \pi m T f_{\textrm{D},1}},& \cdots,&   e^{j2 \pi (M-1) T f_{\textrm{D},1}}  \end{bmatrix} }^\textrm{T}
%& = {\bf k}_{1,\textrm{D}}
%\end{aligned}.
%\end{equation}

\subsubsection{Multiple targets}\label{sec:Delay_Doppler_4_2}

Compared with the single target scenario, it is more difficult to obtain TOs and CFOs for multiple targets by CCCS algorithm because of the problem of inter-target correlation interference. To overcome this challenge, we propose a multi-target alignment algorithm to handle the outputs of the CCCS method for multiple targets.

%描述具体操作步骤
{\color{black} According to \eqref{equ:CSCC_3}, \eqref{equ:CSCC_4}, \eqref{equ:CSCC_5} and \eqref{equ:CSCC_6}, we can obtain the active sensing information vector, ${\bf k}_{1,{R}}(k)$, ${\bf k}_{2,{R}}(k)$ and passive sensing information vector, ${\bf k}_{1,{D}}(k)$, ${\bf k}_{2,{D}}(k)$.}
For mitigating $\Delta \tau_{l} + \delta_\tau(m)$, the CCCS algorithm between ${\bf k}_{1,{R}}(k)$ and ${\bf k}_{2,{R}}(k)$ generates

\begin{equation} \label{equ:CSCC_4_2_1}
\begin{aligned}
{\bf \rho}_{{R}}(k) &=  {\rm{diag}} ({\bf k}_{2,{R}}^{\textrm H} (k)) \cdot {\bf k}_{1,{R}}(k)  &=
{\begin{bmatrix}  \beta_{R} (0,k), &\cdots, &\beta_{R} (n,k),& \cdots,& \beta_{R} (N-1,k)  \end{bmatrix} }^\textrm{T}
%, \in {{\mathcal C}^{N \times 1}}
\end{aligned},
\end{equation}
where
\begin{equation} \label{equ:CSCC_4_2_2}
\begin{aligned}
\beta_{R} (n,k) &= \sum_{l_1=0}^{L-1} \sum_{l_2=0}^{L-1} \alpha_{1,l_1} \alpha_{2,l_2} \cdot e^{j (\Omega_{l_1}-\Omega_{l_2}) k} e^{j 2 \pi n \Delta f (\tau_{2,l_2} - \tau_{1,l_1} + \delta_\tau(m))}
\end{aligned},
\end{equation}
which can be split into two parts, $D_{R} (n,k)$ and $I_{R} (n,k)$.
\begin{equation} \label{equ:CSCC_4_2_3}
\begin{aligned}
D_{R} (n,k)
&= \sum_{l=0}^{L-1} \alpha_{1,l} \alpha_{2,l} e^{j (\Omega_{l}-\Omega_{l}) k} e^{j 2 \pi n \Delta f (\tau_{2,l} - \tau_{1,l} + \delta_\tau(m))} \\
&=\sum_{l=0}^{L-1} \alpha_{1,l} \alpha_{2,l} e^{j 2 \pi n \Delta f (\Delta {\tau}_l + \delta_\tau(m))}
\end{aligned},
\end{equation}
\begin{equation} \label{equ:CSCC_4_2_4}
\begin{aligned}
I_{R} (n,k) &= \sum_{l_1=0}^{L-1} \sum_{l_2=0,l_2 \ne l_1}^{L-1} \alpha_{1,l_1} \alpha_{2,l_2} e^{j (\Omega_{l_1}-\Omega_{l_2}) k} \cdot e^{j 2 \pi n \Delta f (\tau_{2,l_2} - \tau_{1,l_1} + \delta_\tau(m))}
\end{aligned}.
\end{equation}
According to \eqref{equ:CSCC_4_2_2}, \eqref{equ:CSCC_4_2_3} and \eqref{equ:CSCC_4_2_4}, the CCCS outputs for multiple targets not only contains information on TOs and CFOs of the same target{\color {black}, $D_{R} (n,k)$,} but also time delay and Doppler spread correlation information between different targets{\color {black}, $I_{R} (n,k)$,} which will interfere with the subsequent TOs and CFOs mitigation operations.
To overcome this problem, we propose a CCCS based algorithm for multiple targets, including two stages, i.e., the extraction stage and the matching stage.
In the extraction stage, we conduct $\delta_\tau(m) + \Delta \tau_{l}$ and $\Delta \tau + \Delta f_{{D},l}$ extraction without differentiating between different targets, and we match them for different targets in the matching stage.

For above CCCS outputs, we have the proposition that $D_{R} (n,k)$ is invariant with $k$ and $I_{R} (n,k)$ is variant with $k$.
Based on the above proposition, we can firstly obtain the set of phase information in $\beta_{R} (n,k)$ by the DFT algorithm \cite{[OFDM]}. Then, the phase set of $D_{R} (n,k)$ can be extracted by continuously adjusting $k$, observing whether the DFT output of corresponding phases change, as shown in Algorithm \ref{alg:CSCC_1}.

\begin{breakablealgorithm}
	\caption{$\Delta {\tau}_l + \delta_\tau(m)$ extraction method}
	\label{alg:CSCC_1}
	\begin{algorithmic}
		\REQUIRE
		\STATE $\cdot$ The CCCS output based on \eqref{equ:CSCC_4_2_1}, ${\bf \rho}_{{R}}(k)$,
		\STATE $\cdot$ the $n$-th element of ${\bf \rho}_{{R}}(k)$ based on \eqref{equ:CSCC_4_2_2}, $\beta_{R} (n,k)$,
		\STATE $\cdot$ the index of antenna, $k=0$,
		\STATE $\cdot$ the threshold, $\varepsilon_1$.
		\STATE Step 1: Calculate the vector sum of $\beta_{R} (n,k)$ on the $N_t$ antennas to enhance $D_{R} (n,k)$. \\
		\begin{equation} \label{equ:CSCC_4_2_5}
		\begin{aligned}
		D_{R} (n) &= \vec \sum_{k=0}^{N_t-1} D_{R} (n,k)
		&= N_t \sum_{l=0}^{L-1} \alpha_{1,l} \alpha_{2,l} e^{j 2 \pi n \Delta f (\Delta {\tau}_l + \delta_\tau(m))}
		\end{aligned},
		\end{equation}
		\begin{equation} \label{equ:CSCC_4_2_6}
		\begin{aligned}
		I_{R} (n) &= \vec \sum_{k=0}^{N_t} I_{R} (n,k)
		&= \sum_{l_1=0}^{L-1} \sum_{l_2=0,l_2 \ne l_1}^{L-1} \vec \sum_{k=0}^{N_t-1} e^{j (\Omega_{l_2} - \Omega_{l_1}) k } \alpha_{1,l_1} \alpha_{2,l_2}
		& e^{j 2 \pi n \Delta f ( \tau_{2,l_2} - \tau_{1,l_1} + \delta_\tau(m))}
		\end{aligned}.
		\end{equation} \\
		\STATE $Proof $: The proof is provided in Appendix \ref{app:B}.
		\STATE Step 2: \\
		2a) Perform the DFT operation on ${\bf \rho}_{{R}}(k)$
		\begin{equation} \label{equ:CSCC_4_2_7}
			\begin{aligned}
				r(q_k) &= {\mathsf {DFT}} \left({\bf \rho}_{{R}}(k)\right) =\sum_{n=0}^{N-1} {\bf \rho}_{{R}}(k) e^{-j \frac{2 \pi}{N} n q_k \Delta f}
			\end{aligned}.
		\end{equation} \\
		2b) Generate $P_k$, which is the set of the delay information, $q_{k,i}$, and the real part of the DFT output, ${\color{black} \rm{real}}(r(q_{k,i}))$.
		\begin{equation} \label{equ:CSCC_4_2_8}
		\begin{aligned}
		P_k &= {\mathcal {S}} { \left(q_{k,i}, { \rm{real}}(r(q_{k,i})) \right),i=0,1,\cdots,L^2}
		\end{aligned},
		\end{equation} \\
		\begin{equation} \label{equ:CSCC_4_2_9}
			\begin{aligned}
				q_{k,i} &= \left \lfloor N \Delta f (\tau_{2,l_2} - \tau_{1,l_1} + \delta_\tau(m)) \right \rfloor
			\end{aligned},
		\end{equation} \\
		where ${\mathcal S}(x_i)$ denotes the set of $x_i$, $q_{k,i}$ denotes the index of peak of DFT outputs, $r(q_{k,i})$ denotes the value of peak of DFT outputs.
		\STATE Step 3: Extract $\Delta {\tau}_l + \delta_\tau(m)$ by adjust $k$, where $\Delta {\tau}_l = \tau_{2,l} - \tau_{1,l}$.
		\STATE 3a) Generate $P_0 = {\mathcal {S}} {(q_{0,i}, { \rm{real}}(r(q_{0,i})))}$ based on step 2.
		\WHILE {$ k \le N_t$ and $P_0$ is updated}
		\STATE $k = k+1$.
		\STATE 3b) Generate $P_k$ based on step 2.
		\FORALL {$(q_{0,i}, {\rm{real}}(r(q_{0,i})))$ in $P_0$}
			\IF {$\left|{\rm{real}}(r(q_{0,i})) - {\rm{real}}(r(q_{k,i}))\right| \ge \varepsilon_1$}
			\STATE 3c) Remove $(q_{0,i}, {\rm{real}}(r(q_{0,i})))$ from $P_0$.
			\ENDIF
		\ENDFOR
		\ENDWHILE
		\ENSURE $P_0$.
	\end{algorithmic}
\end{breakablealgorithm}
So far, $\Delta \tau_{l} + \delta_\tau(m)$ has been estimated, which can be expressed as
\begin{equation} \label{equ:CSCC_4_2_10}
\begin{aligned}
P_{{g}} &= {\mathcal S}(\Delta \tau_{0} + \delta_\tau(m), \cdots, \Delta \tau_{l} + \delta_\tau(m), \cdots, \Delta \tau_{L-1} + \delta_\tau(m))
\end{aligned}.
\end{equation}
The delay estimation of active and passive sensing can be obtained by DFT \cite{[OFDM]}
\begin{equation} \label{equ:CSCC_4_2_11}
\begin{aligned}
P_{m} &= {\mathcal S}(\tau_{1,0}, \cdots, \tau_{1,l}, \cdots, \tau_{1,L-1}) \\
P_{b} &= {\mathcal S}(\tau_{2,0} + \delta_\tau(m), \cdots, \tau_{2,l} + \delta_\tau(m), \cdots, \tau_{2,L-1} + \delta_\tau(m))
\end{aligned}.
\end{equation}
Although $\Delta \tau_{l} + \delta_\tau(m)$ has been estimated in $P_{g}$, there is a mismatch between $P_{g}$ and $P_{m}$ and $P_{b}$. Recognizing this fact, we propose a matching method to achieve the matching of active and passive delay estimation of different targets, as shown in Algorithm \ref{alg:CSCC_2}.
\begin{breakablealgorithm}
	\caption{Matching method for delay estimation of different targets}
	\label{alg:CSCC_2}
	\begin{algorithmic}
		\REQUIRE
		\STATE $\cdot$ $P_{b}$, $P_{m}$, $P_{g}$, \\
		$\cdot$ the threshold, $\varepsilon_2$, \\
		$\cdot$ the empty matching set, $C_{R} = {\mathcal S}(\cdot)$.
		\FOR {$P_{g}(i)$ in $P_{g}$ and $P_{m}(i)$ in $P_{m}$ and $P_{b}(i)$ in $P_{b}$}
			\IF {$|P_{m}(i) + P_{g}(i) - P_{b}(i)| \le \varepsilon_2$}
				\STATE Add $(P_{m}(i),P_{g}(i),P_{b}(i))$ to
				$C_{R}$.
			\ENDIF
		\ENDFOR
		\ENSURE $C_{R}$.
	\end{algorithmic}
\end{breakablealgorithm}
%So far, TOs $\delta_\tau(m)$ and deviations in the time delay $\Delta \tau$ for active and passive of multiple targets have been mitigated.
As for mitigating $\Delta f_{{D},l} + \delta_f(m)$, the outputs of the CCCS algorithm between ${\bf k}_{1,{D}}(k)$ and ${\bf k}_{2,{D}}(k)$ can be expressed as
\begin{equation} \label{equ:CSCC_4_2_13}
\begin{aligned}
{\bf \rho}_{{V}}(k) &=  {\color{black} {\rm{diag}}} ({\bf k}_{2,{D}}^{\textrm H} (k)) \cdot {\bf k}_{1,{D}}(k)  \\
&=
{\begin{bmatrix}  \beta_{V} (0,k), &\cdots, &\beta_{V} (m,k),&\cdots,&\beta_{V} (M-1,k)  \end{bmatrix} }^\textrm{T}
%, \in {{\mathcal C}^{N \times 1}}
\end{aligned},
\end{equation}
where
\begin{equation} \label{equ:CSCC_4_2_14}
\begin{aligned}
\beta_{V} (m,k) &= \sum_{l_1=0}^{L-1} \sum_{l_2=0}^{L-1} \alpha_{1,l_1} \alpha_{2,l_2} e^{j (\Omega_{l_1}-\Omega_{l_2}) k} \cdot e^{j 2 \pi m T (f_{{D},2,l_2} - f_{{D},1,l_1} + \delta_f(m))}
\end{aligned},
\end{equation}
which can be split into two parts, $D_{V} (m,k)$ and $I_{V} (m,k)$.
\begin{equation} \label{equ:CSCC_4_2_15}
\begin{aligned}
D_{V} (m,k) &= \sum_{l_1=0}^{L-1} \sum_{l_2=0,l_2=l_1}^{L-1} \alpha_{1,l_1} \alpha_{2,l_2} e^{j (\Omega_{l_1}-\Omega_{l_2}) k} \cdot e^{j 2 \pi m T (f_{{D},2,l_2} - f_{{D},1,l_1} + \delta_f(m))} \\
&= \sum_{l=0}^{L-1} \alpha_{1,l} \alpha_{2,l} e^{j (\Omega_{l}-\Omega_{l}) k} e^{j 2 \pi m T (\Delta f_{{D},l} + \delta_\tau(m))} \\
&=\sum_{l=0}^{L-1} \alpha_{1,l} \alpha_{2,l} e^{j 2 \pi m T (\Delta f_{{D},l} + \delta_\tau(m))}
\end{aligned},
\end{equation}
\begin{equation} \label{equ:CSCC_4_2_16}
\begin{aligned}
I_{V} (m,k) &= \sum_{l_1=0}^{L-1} \sum_{l_2=0,l_2 \ne l_1}^{L-1} \alpha_{1,l_1} \alpha_{2,l_2} e^{j (\Omega_{l_1}-\Omega_{l_2}) k} \cdot e^{j 2 \pi m T (f_{{D},2,l_2} - f_{{D},1,l_1} + \delta_f(m))}
\end{aligned}.
\end{equation}
Similarly, $D_{V} (m)$ can be obtained by an operation like \eqref{equ:CSCC_4_2_5}.
Most operation of CCCS for mitigating $\Delta f_{\textrm{D},l} + \delta_f(m)$ is similar to that for mitigating $\Delta f_{{D},l} + \delta_f(m)$, which has been introduced above. The main difference is how to extract the Doppler information from $D_{V} (m)$.
%, as shown in Step 2 of Algorithm \ref{alg:CSCC_1}.

For case 1: CFO is almost constant within the sensing signal containing $M$ OFDM symbols,
\begin{equation} \label{equ:CSCC_4_2_17}
\begin{aligned}
\delta_f(0) \approx,\cdots,\approx \delta_f(m)\approx,\cdots,\approx \delta_f(M-1)
\end{aligned}.
\end{equation}
Then, $\Delta f_{{D},l} + \delta_f(m)$ is invariant with $m$, which means that the Doppler information from $D_{V} (m)$ can be obtained by DFT.

For case 2: CFO is not a constant within the sensing signal containing $M$ OFDM symbols,
\begin{equation} \label{equ:CSCC_4_2_18}
\begin{aligned}
\delta_f(0) \ne,\cdots,\ne \delta_f(m)=,\cdots,\ne \delta_f(M-1)
\end{aligned}.
\end{equation}
Then, $\Delta f_{{D},l} + \delta_f(m)$ is variant with $m$, which means that the Doppler information from $D_{V} (m)$ cannot be obtained by the DFT algorithm.
Therefore, the performance of CCCS for mitigating $\Delta f_{{D},l} + \delta_f(m)$ is affected by the variance of CFO, which will be analyzed and simulated in detail in Section \ref{sec:Simulation}.

\subsection{Delay and Doppler spread Estimation by two dimensional (2D) DFT }\label{sec:Delay_Doppler_5}

As mentioned in Section \ref{sec:Delay_Doppler_4}, after mitigating $\Delta \tau_{l} + \delta_\tau(m)$ and $\Delta f_{{D},l} + \delta_f(m)$, the active and passive sensing information vector can be translated into a linear phase shift in the modulation symbols and the carrier frequency domain.
%the passive sensing information vector after mitigating $\Delta \tau_{l} + \delta_\tau(m)$ and $\Delta f_{{D},l} + \delta_f(m)$ can be represented as ${\bf k}_{2,{R}}^{'}$ and ${\bf k}_{2,{D}}^{'}$.
%Then, the active and passive sensing fusion information vectors can be expressed as ${\bf k}_{1,\textrm{R}} + {\bf k}_{2,\textrm{R}}^{'}$ and ${\bf k}_{1,\textrm{D}} + {\bf k}_{2,\textrm{D}}^{'}$, where $\tau_{1,l}$ and $f_{\textrm{D},1,l}$ translate into a linear phase shift in the modulation symbols and the carrier frequency domain.
The most convenient way to evaluate the range to the targets, is to compute inverse DFT (IDFT) of the range steer vector
\begin{equation} \label{equ:CSCC_5_1}
\begin{aligned}
r(i)
%&= {\mathcal {IDFT}}({\bf k}_{1,\textrm{R}} + {\bf k}_{2,\textrm{R}}^{'}) \\
&= \frac{1}{N_I} \sum_{n=0}^{N_I-1} \sum_{l=0}^{L-1} (\alpha_{1,l} + \alpha_{2,l}) {e^{-j2 \pi n \Delta f (\tau_{1,l}) } e^{j\frac{2 \pi}{N_I} n i}}, \quad i = 0, 1, \cdots, N_I-1
\end{aligned},
\end{equation}
{\color{black}
where $N_I$ is the number of IDFT points.}
It can be seen that, the two exponential terms in \eqref{equ:CSCC_5_1} cancel each other and result in unity, under the condition
\begin{equation} \label{equ:CSCC_5_2}
\begin{aligned}
& i = \left \lfloor \Delta f N_I \tau_{1,l} \right \rfloor
\end{aligned},
\end{equation}
where $\left \lfloor \cdot \right \rfloor$ denotes the operation of round down. It means that a peak will occur at this index of $i$ in the time response $r(i)$.
Similarly, the estimation of Doppler spread can be solved by applying DFT to the Doppler spread steer vector
%${\bf k}_{1,\textrm{D}} + {\bf k}_{2,\textrm{D}}^{'}$
\begin{equation} \label{equ:CSCC_5_3}
\begin{aligned}
v(i)
%&= {\mathsf {DFT}}({\bf k}_{1,\textrm{D}} + {\bf k}_{2,\textrm{D}}^{'}) \\
&=  \sum_{m=0}^{M_D-1} \sum_{l=0}^{L-1} (\alpha_{1,l} + \alpha_{2,l}) {e^{j2 \pi m T  f_{\textrm{D},1,l} } e^{-j\frac{2 \pi}{M_D} m i}}, \quad i = 0, 1, \cdots, M_D-1
\end{aligned},
\end{equation}
{\color{black}
where $M_D$ is the number of DFT points.}
Cancellation of the exponential terms, and constructive superposition, results for the index
\begin{equation} \label{equ:CSCC_5_4}
\begin{aligned}
& i = \left \lfloor T M_D f_{\textrm{D},1,l} \right \rfloor
\end{aligned}.
\end{equation}
Then, delay and Doppler spread of the $l$-th target can be estimated based on \eqref{equ:CSCC_5_3} and \eqref{equ:CSCC_5_4}.
Further, ranging and velocity measurement can be realized by \eqref{equ:CSCC_7} and \eqref{equ:CSCC_9}.

\subsection{Complexity Analysis}\label{sec:Delay_Doppler_6}
\begin{table}[ht]
	\caption{Complexity of Delay and Doppler spread estimation}
	\centering
	\label{label:Complexity_1}
	\begin{tabular}{l|c|c|c|c|c}
		\hline \hline
		Operation & CCCS & Algorithm 1 & Algorithm 2 & 2D DFT & Overall \\ \hline
		Delay estimation & $\mathcal{O}(N^2 \times N_t)$ &  $Nlog(N) \times N_t$ &  $\mathcal{O}(L^3)$ &  $\mathcal{O}(Nlog(N))$ & $\mathcal{O}(N^2 \times N_t + L^3)$ \\ \hline
		Doppler spread estimation & $\mathcal{O}(M^2 \times M_t)$ & $Mlog(M) \times N_t$ & $\mathcal{O}(L^3)$ & $\mathcal{O}(Mlog(M))$ & $\mathcal{O}(M^2 \times N_t + L^3)$ \\ \hline
	\end{tabular}
\end{table}
In this subsection, we analyze the computational complexity of cooperative sensing for delay and Doppler spread estimation.
Since the delay and Doppler spread are estimated in parallel, without loss of generality, we only analyze the complexity of the delay estimation.
There are four main computations, CCCS, Algorithm \ref{alg:CSCC_1}, Algorithm \ref{alg:CSCC_2} and delay estimation by DFT.
Since the size of ${\bf k}_{1,{R}}(k)$ is $1 \times N$ and the size of ${\rm{diag}}({\bf k}_{2,{R}}(k))$ is $N \times N$, the complexities of CCCS is $\mathcal{O}(N^2)$. For $N_t$ antennas, the complexities of CCCS is $\mathcal{O}(N^2 \times N_t)$.
In Algorithm \ref{alg:CSCC_1}, {the complexity of calculating the vector sum of $\beta_{R}(n,k)$ with the dimension of $1 \times 1$ on $N_t$ antennas is $\mathcal{O}(N_t)$, the complexity of step 2 is $\mathcal{O}(Nlog(N))$, for $N_t$ antennas, the complexity is updated to $Nlog(N) \times N_t$.}
In Algorithm \ref{alg:CSCC_2}, since the size of $P_{m}$, $P_{b}$ and $P_{g}$ are all close to $1 \times L$, the complexities of matching method is $\mathcal{O}(L^3)$.
According to \cite{[OFDM]}, the complexities of DFT is $\mathcal{O}(Nlog(N))$.
Therefore, the overall complexity for estimating delay is $\mathcal{O} (N^2 \times N_t + L^3)$
Likewise, the overall complexity for estimating Doppler spread can be derived as $\mathcal{O} (M^2 \times N_t + L^3)$.
The details of complexity analysis are summarized in Table \ref{label:Complexity_1}

\section{AoA Estimation based on cooperative sensing}\label{sec:AOA}
{\color{black}
In this section, we propose a low complexity AoA estimation algorithm based on cooperative sensing, which contains two stages, i.e., the coarse AoA estimation (CAE) stage and the fine AoA estimation (FAE) stage.
%
%When the antenna scale is large, the number of spatial samples that SBS can obtain at a time is large enough to estimate AoA directly based on the spatial domain by one-shot measurement, which means that AoA estimation can be done at the same time as the Doppler spread and delay estimation \cite{[AOA_41]}. However, as the number of antennas increases, the computational complexity and energy consumption also increase, making it more difficult to implement the hardware. Recognizing this fact, Ni {\it et al.} proposed a high-accuracy AoA estimation method by integrating both the time-domain and the frequency-domain signals into the spatial domain \cite{[UL_sensing]}. However, the algorithm enlarges the length of the spatial array response vectors by integrating both
%the time-domain and the frequency-domain signals into the spatial domain, which in turn improves the accuracy of AoA estimation, but the computational complexity increases as well.
%To compensate for the disadvantage mentioned above, we propose a low complexity AoA estimation algorithm based on DFT and TSF-MUSIC to achieve high accuracy and low complexity AoA estimation when the antenna scale is not necessarily very large.
%The algorithm contains two stages, i.e., the coarse AoA estimation (CAE) stage and the fine AoA estimation (FAE) stage.
%
In the CAE stage, we obtain a rough AoA estimation of targets by cooperative active and passive sensing.
In the FAE stage, we can obtain a fine AoA of targets by FRFT algorithm within the range of rough AoA estimation.
\subsection{CAE by Cooperative Sensing}\label{sec:AOA_1_1}
\begin{figure}[ht]
	\includegraphics[scale=0.45]{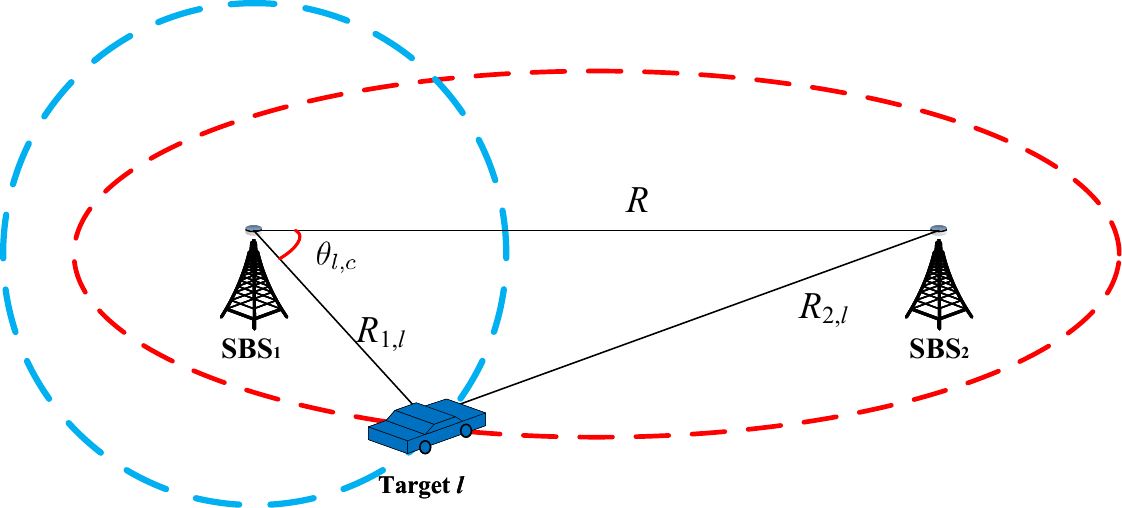}
	\centering
	\caption{CAE by cooperative sensing.}
	\label{fig:AOA_CP}
\end{figure}
As Fig. \ref{fig:AOA_CP} shows, the distance between $\rm {SBS}_1$ and $\rm {SBS}_2$ is $R$, the ranging result of active sensing is $R_{1,l}$, the ranging result of passive sensing is $R_{1,l}+R_{2,l}$. Then the rough AoA estimated result of the $l$-th target, $\Omega_{l,c}$ can be expressed as
\begin{equation} \label{equ:AOA_CP_1}
	\begin{aligned}
		\theta_{l,c} &= {\rm acos} \left(\frac{R_{1,l}^2 + R^2 - R_{2,l}^2}{2R R_{1,l}} \right) \\
		\Omega_{l,c} &= \frac{2 \pi d}{\lambda} {\rm cos} \left(\theta_{l,c}\right) = \frac{2 \pi d}{\lambda} \cdot \frac{R_{1,l}^2 + R^2 - R_{2,l}^2}{2R R_{1,l}}
	\end{aligned},
\end{equation}

%%设置角度范围
\subsection{FAE by FRFT} \label{sec:AOA_1_2}
%Since the ranging results of active and passive sensing are not
The accuracy of $\Omega_{l,c}$ is not high because there are errors in the ranging results of active and passive sensing.
To obtain high accuracy AoA estimation, we adopt the FRFT algorithm \cite{[QHY]} for the received steer vector of targets
%, which has been introduced in \eqref{equ:Received_Signal_Model_1}.
%\begin{equation} \label{equ:AOA_1_1_1}
%\begin{aligned}
%{\bf a}(\Omega_l) &= \sum_{l=0}^{L-1}
%{\begin{bmatrix}  1,& e^{j\Omega_l},& \cdots,& e^{j(N_t-1)\Omega_l}  \end{bmatrix} }^\textrm{T}
%%, \in {{\mathcal C}^{N_t \times 1}}
%\end{aligned},
%\end{equation}
%where $\Omega_l = \frac{2 \pi d}{\lambda} {\rm{cos}}(\theta_l)$.
%To estimate $\Omega_l$, we can compute the FRFT of ${\bf a}(\Omega_l)$
\begin{equation} \label{equ:AOA_CP_2}
	\begin{aligned}
		{\mathsf{AoA}}(i) &= \sum_{k=0}^{N_t N_f-1} \sum_{l=0}^{L-1} e^{j\Omega_l k} e^{-j \frac{2 \pi}{N_t N_f} ki}, \quad i = 0,1,...,N_tN_f
	\end{aligned},
\end{equation}
where $N_f$ is the fractional index of FRFT.
However, the complexity of FRFT is $\mathcal{O} (N_t^2 N_f^2)$, which gets larger squared as $N_f$ gets larger.
To decrease the complexity of FRFT, we adopt the FRFT algorithm within the range of $\left[\Omega_{l,c}-\Omega, \Omega_{l,c} + \Omega\right]$
\begin{equation} \label{equ:AOA_CP_3}
	\begin{aligned}
		{\mathsf{AoA}}(i) &= \sum_{k=0}^{N_t N_f-1} \sum_{l=0}^{L-1} e^{j\Omega_l k} e^{-j \frac{2 \pi}{N_t N_f} ki}, \quad i = i_s,...,i_e
	\end{aligned},
\end{equation}
\begin{equation} \label{equ:AOA_CP_4}
	\begin{aligned}
		i_s &= \left \lfloor \frac{N_t N_f (\Omega_{l,c}-\Omega)}{2 \pi} \right \rfloor  \\
		i_e &= \left \lfloor \frac{N_t N_f (\Omega_{l,c}+\Omega)}{2 \pi} \right \rfloor
	\end{aligned},
\end{equation}
where $\Omega$ is the uncertainty range of $\Omega_{l,c}$.
Then, the fine AoA estimated result, $\Omega_{l,f}$ can be derived as
\begin{equation} \label{equ:AOA_CP_5}
\begin{aligned}
&\Omega_{l,f} \in \left[\frac{2 \pi i_l}{N_t N_f}, \frac{2 \pi i_l}{N_t N_f} + 1 \right) \\
&\theta_{l,f} \in \left[ {\color{black}{\rm{acos}}} \left(\frac{\lambda i_l}{d N_t N_f} \right), {\rm{acos}} \left(\frac{\lambda i_l}{d N_t N_f }+\frac{\lambda}{2 \pi d}    \right )  \right)
\end{aligned},
\end{equation}
where $i_l$ is the index of the peak of ${\mathsf{AoA}}(i)$.

\begin{breakablealgorithm}
	\caption{AoA estimation based on cooperative sensing}
	\label{alg:FRFT}
	\begin{algorithmic}
		\REQUIRE
		The waveform length $\lambda$, the distance between two adjacent antenna elements $d$, the received steer vector $\bf a (\Omega_{l})$, the ranging results of active and passive sensing $R_{1,l}$, $R_{1,l} + R_{2,l}$, the uncertainty range, $\Omega$;
		\STATE 1) CAE by cooperative sensing:
		\STATE \quad 1a) Obtain coarse AoA estimated result, $\Omega_{l,c}$ according to \eqref{equ:AOA_CP_1}.
		\STATE 2) FAE by FRFT:
		\STATE \quad 2a) Determine the range of FRFT based on \eqref{equ:AOA_CP_4}.
		\STATE \quad 2b) Obtain the fine AoA estimated result, $\Omega_{l,c}$ according to \eqref{equ:AOA_CP_3} and \eqref{equ:AOA_CP_5}.
		\ENSURE $\Omega_{l,f}$.
	\end{algorithmic}
\end{breakablealgorithm}

\subsection{Complexity Analysis}\label{sec:AOA_2}

In this subsection, we analyze the computational complexity of Algorithm \ref{alg:FRFT}, which includes two parts.
In the CAE stage, the complexity of obtaining $\Omega_{l,c}$ is $\mathcal{O} (1)$.
In the FAE stage, the complexity of FRFT based on the rough AoA estimation is $\mathcal{O} \left(\frac{N_t^2 N_f^2 \Omega}{\pi} \right)$.
Therefore, the overall computational complexity of Algorithm \ref{alg:FRFT} is $\mathcal{O} \left (\frac{N_t^2 N_f^2 \Omega}{\pi} \right)$, which is $\frac{\Omega}{\pi}$ times less complex than the traditional FRFT algorithm. The complexities of the main steps of AoA estimation are summarized in Table \ref{label:Complexity_2} and are compared with those of the traditional FRFT method.

%Similar to the delay estimation, the complexity of CAE by DFT is $\mathcal{O} (N_t log(N_t))$.
%For FAE by TSF-MUSIC, one main computation is the SVD of ${\bf S}$ with the dimension of $C N_t \times 2 C_1$. Since we only need the left singular matrix ${\bf U}_s$, the complexity is $\mathcal{O} (2C^2 N_t^2 C_1)$.
%Another main computation is obtaining the AoA estimation from MUSIC spatial spectrum in \eqref{equ:AOA_1_2_4}.
%With the help of CAE, the testing AoA range is the coarse angle range, which means that the size of ${\bf a}_\textrm{tr}(\Omega)$ is $1 \times C_t$. Then, the complexity of AoA estimation from MUSIC spatial spectrum is $\mathcal{O} (2C_t  C N_t C_1) + \mathcal{O}( 2 C_t C_1) \approx \mathcal{O} (2C_t C N_t C_1)$.
%Other steps in MUSIC have much lower complexity and can be omitted.
%Therefore, the overall complexity for estimating AoA is $\mathcal{O} (N_t log(N_t)) + \mathcal{O} (2C^2 N_t^2 C_1) + \mathcal{O} (2C_t C N_t C_1) \approx \mathcal{O} (2C^2 N_t^2 C_1) + \mathcal{O} (2C_t C N_t C_1)$. If $C_t \le C N_t$, the overall complexity can be expressed as $\mathcal{O} (2C^2 N_t^2 C_1)$.
%For conventional MUSIC without the help of CAE, the testing AoA range is the whole angle range, which means that the size of ${\bf a}_\textrm{tr}(\Omega)$ is $1 \times C_t N_t$.
%The complexity of AoA estimation from MUSIC spatial spectrum is $\mathcal{O} (2C_t  C N_t^2 C_1) + \mathcal{O}( 2 C_t N_t C_1) \approx \mathcal{O} (2C_t C N_t^2 C_1)$.
\begin{table}[h]
	\caption{Complexity of AoA estimation}
	\centering
	\label{label:Complexity_2}
	\begin{tabular}{l|c|c|c|c}
		\hline \hline
		Operation & CAE based on cooperative sensing & FRFT with CAE & FRFT without CAE & Overall \\ \hline
		Algorithm \ref{alg:FRFT} & $\mathcal{O}(1)$ & $\mathcal{O} \left(\frac{N_t^2 N_f^2 \Omega}{\pi}\right)$ &   & $\mathcal{O} \left(\frac{N_t^2 N_f^2 \Omega}{\pi}\right)$ \\ \hline
		Traditional FRFT & & & $\mathcal{O} \left(N_t^2 N_f^2\right)$  & $\mathcal{O} \left(N_t^2 N_f^2\right)$ \\ \hline
	\end{tabular}
\end{table}

%The complexities of the main steps of AOA estimation are summarized in Table \ref{label:Complexity_2} and are compared
%with those of the traditional FRFT method. From Table \ref{label:Complexity_2}, we
%note that the overall complexity of Algorithm \ref{alg:FRFT} is lower than
%TSF-MUSIC, which is because TSF-MUSIC requires $N_t$ times the number of candidates.

}
\section{Simulation Results}\label{sec:Simulation}

In this section, we first analyze and simulate the accuracy of range and velocity estimation of passive sensing with TOs and CFOs, and then analyze the performance improvement in delay, Doppler spread and AoA estimation.
Simulation parameters used in this section are shown in table \ref{Parameter:simulation} \cite{[UL_sensing],[5G_signal],[TO_CFO_1]}.

\begin{table}[h]
	\caption{Simulation parameters adopted in this paper.}
	\centering
	\label{Parameter:simulation}
	\begin{tabular}{l|l|l|l|l|l}
		\hline
		\hline
		Items & Value & Meaning of the parameter & Items & Value & Meaning of the parameter \\ \hline
		$f_\textrm{c1}$ & 4 GHz \cite{[5G_signal]} & Carrier frequency of active sensing & $f_\textrm{c2}$ & 4.2 GHz \cite{[5G_signal]} & Carrier frequency of passive sensing \\ \hline
		$c$ & $3\cdot10^8$ m/s & Speed of light & $\Delta f$ & $120$ kHz \cite{[5G_signal]} & Carrier frequency \\ \hline
		$M$ & 256 \cite{[UL_sensing]} & Number of OFDM symbols & $N$ & 1024 \cite{[5G_signal]} & Number of subcarriers \\ \hline
		\color{black}
		$M_D$ & \color{black} 2560 & \color{black} Number of DFT points & \color{black} $N_I$ & \color{black} 10240 & \color{black} Number of IDFT points \\ \hline
%		$T_p$ & 8.3 us \cite{[5G_signal]} & OFDM symbol period & $T_c$ & 2.08 us \cite{[5G_signal]} & CP period \\ \hline
		$T$ & 10.38 us \cite{[5G_signal]} & The whole OFDM period & $B$ & 123 MHz \cite{[5G_signal]} & Frequency bandwidth \\ \hline
		$N_t$ & [4,8] \cite{[UL_sensing]} & Number of antenna array & $L$ & [3,5] & Number of targets \\ \hline
		$R_l$ & [70,100,130]m & Range of targets & $V_l$ & [15,25,35]m/s & Velocity of targets \\ \hline
		$\theta_l$ & $[25,30,35]^o$ & AoA of targets & $\rm SNR_m$ & [-30,30] dB & SNR of active sensing received signals \\ \hline
		$\rm SNR_b$ & [-30,30] dB & SNR of passive sensing received signals & $\rm SNR_r$ & $\frac{\rm SNR_b}{\rm SNR_m}$ & Power ratio between two types of sensing \\ \hline
		$E(\delta_\tau)$ & [10,1000] ns & Mean of TOs & $E(\delta_f) $ & [0.01,0.2] $\Delta f$ & Mean of CFOs \\ \hline
		$ V(\delta_\tau)$ & [1,100] ns & Variance of TOs & $ V(\delta_f)$ & [0.01,0.2] $\Delta f$ & Variance of CFOs \\ \hline
		$ \varepsilon_1$ & 0.01 & Threshold in Algorithm \ref{alg:CSCC_1} & $ \varepsilon_2$ & 0.01 & Threshold in Algorithm \ref{alg:CSCC_2} \\ \hline
	\end{tabular}
\end{table}

\subsection{Range and Velocity Estimation of Passive Sensing with TOs and CFOs}\label{sec:Simulation_1}

\begin{figure*}[ht]
	\centering
	\subfigure[\scriptsize{NMSE of range with TOs}.]{\includegraphics[width=0.45\textwidth]{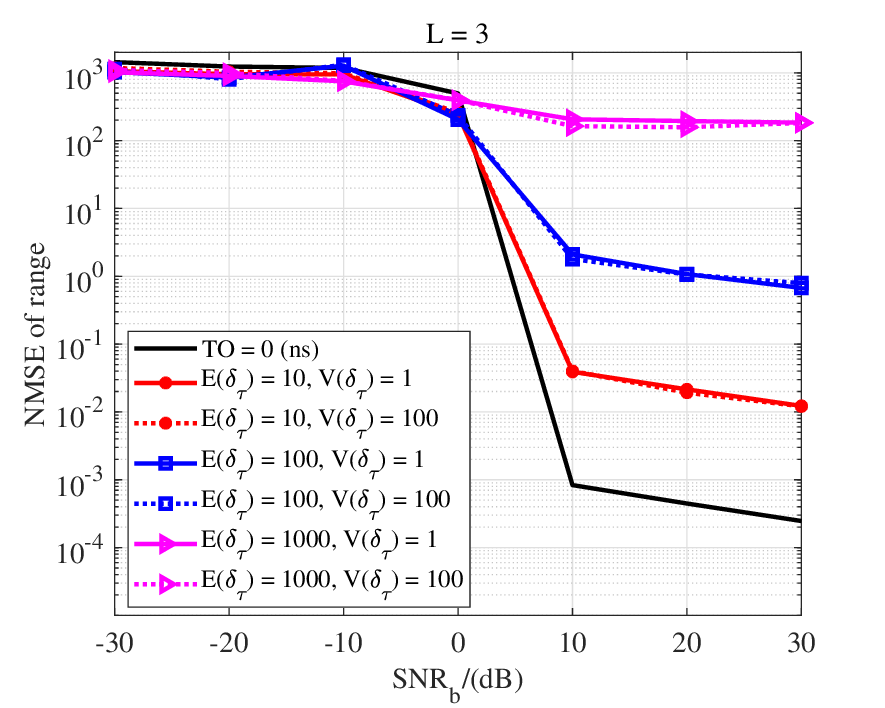}%
		\label{fig:TO_R}}
	\hfil
	\subfigure[\scriptsize{NMSE of velocity with TOs}.]{\includegraphics[width=0.45\textwidth]{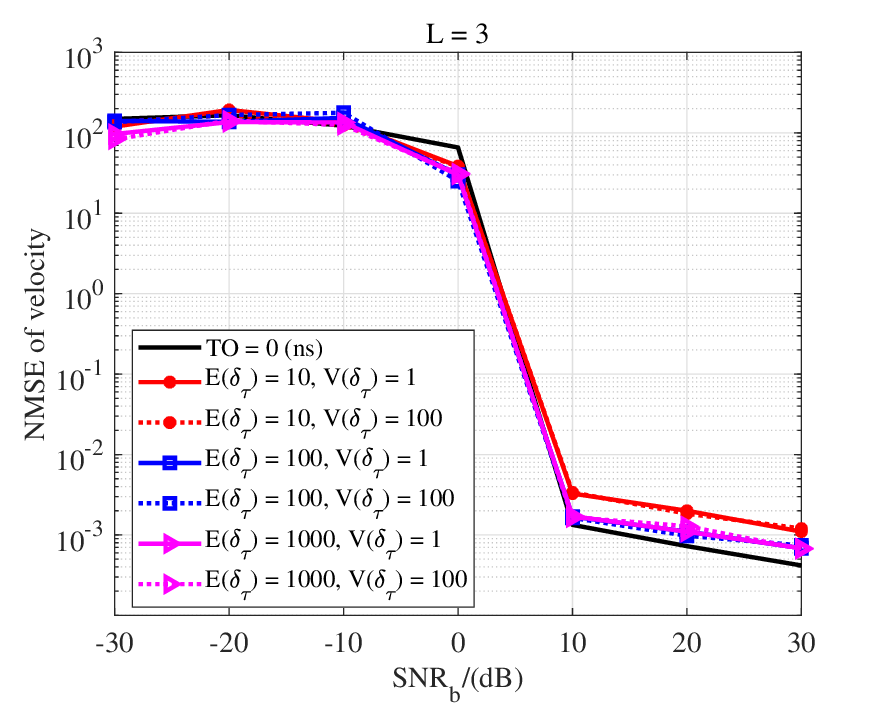}%
		\label{fig:TO_V}}
	\hfil
	\subfigure[\scriptsize{NMSE of range with CFOs}.]{\includegraphics[width=0.45\textwidth]{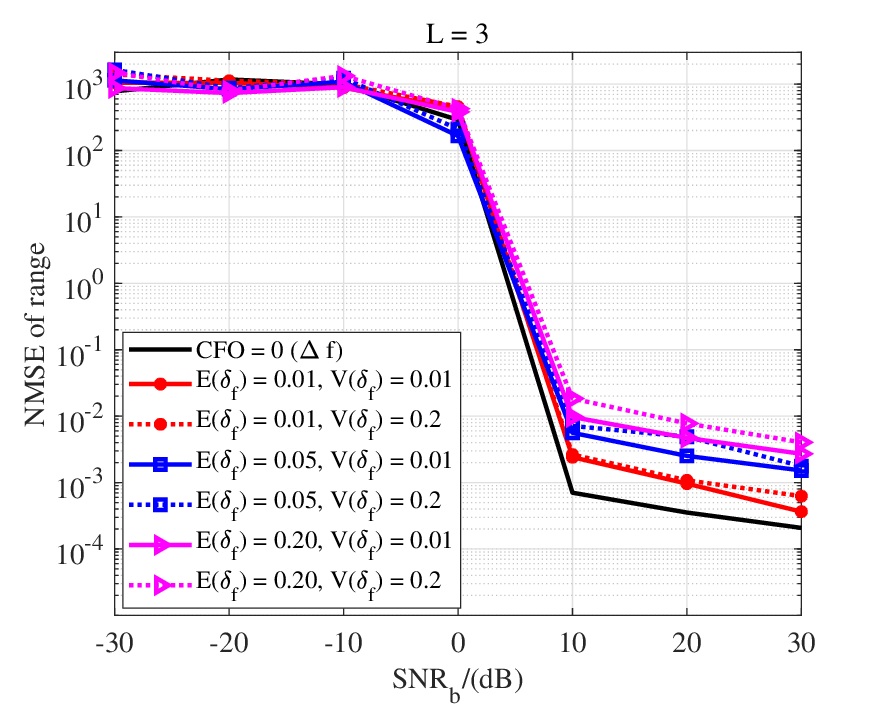}%
		\label{fig:CFO_R}}
	\hfil
	\subfigure[\scriptsize{NMSE of velocity with CFOs}.]{\includegraphics[width=0.45\textwidth]{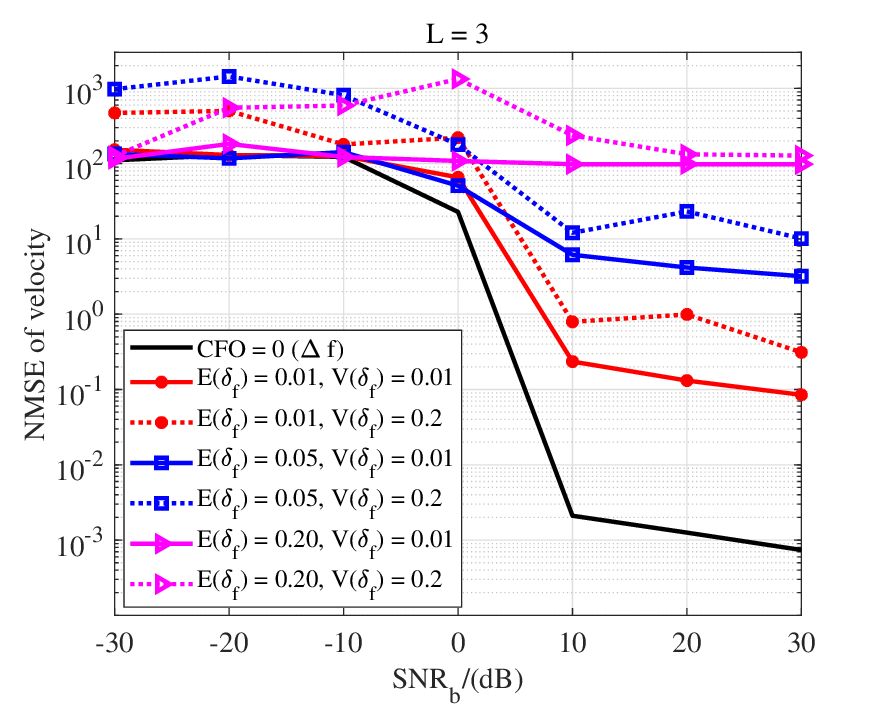}%
		\label{fig:CFO_V}}
	\hfil
	\caption{NMSE of range and velocity with TOs and CFOs.}
	\label{fig:TO_CFO_R_V}
\end{figure*}

In this section, we analyze and simulate the performance of passive sensing with TOs and CFOs.
Fig. \ref{fig:TO_CFO_R_V} presents the normalized mean-squared-error (NMSE) for range and velocity, which can be expressed as
\begin{equation} \label{equ:Simulation_1_1}
\begin{aligned}
\Gamma_{R} &= \frac{1}{L} \sum_{l=0}^{L-1} \frac{ \left|R_{{e},l} - R_{r,l}\right|^2}{R_{r,l}^2}, \qquad
\Gamma_{V} &= \frac{1}{L} \sum_{l=0}^{L-1} \frac{\left|V_{{e},l} - V_{r,l}\right|^2}{V_{r,l}^2}
\end{aligned},
\end{equation}
where $R_{r,l}$ and $V_{r,l}$ are the actual range and velocity of targets, $R_{{e},l}$ and $V_{{e},l}$ are the estimated range and velocity.
As Fig. \ref{fig:TO_R} and Fig. \ref{fig:TO_V} show that the effect of TOs on velocity estimation is greater than that no range estimation, which is consistent with the theoretical analysis.
Under the condition of low $\rm SNR_b$, NMSE of range $\Gamma_{R}$ is large and the performance of range estimation is poor, and the larger the mean value of TOs $E(\delta_\tau)$ is, the larger the fluctuation of $\Gamma_{R}$ is.
As $\rm SNR_b$ is raised, $\Gamma_{R}$ is stable. In the meanwhile, the larger $E(\delta_\tau)$ is, the larger $\Gamma_{R}$ is, leading to poorer ranging performance.
Moreover, by comparing the solid line and the dashed line in Fig. \ref{fig:TO_R}, it can be found that the solid line is close to the dashed line under the condition of the same $E(\delta_\tau)$, which means that the variance of TOs $V(\delta_\tau)$ does not have a great influence on the range estimation.

Fig. \ref{fig:CFO_R} and Fig. \ref{fig:CFO_V} present the NMSE of range and velocity with TOs and CFOs.
According to Fig. \ref{fig:CFO_R}, $\Gamma_{R}$ decreases as $\rm SNR_b$ increases and gradually stabilizes, and CFOs have little effect on the stable $\Gamma_{R}$.
However, when the mean value of CFOs $E(\delta_f)$ is $0.2 \Delta f$, $\rm SNR_b$ required for $\Gamma_{R}$ stabilization is larger due to the ICI caused by CFOs, which in turn affects the range estimation at low $\rm SNR_b$ conditions.
According to the solid line in Fig. \ref{fig:CFO_V}, under the condition of low variance of CFOs $V(\delta_f)$, NMSE of velocity $\Gamma_{V}$ will gradually stabilizes with high $\rm SNR_b$, but $\Gamma_{V}$ increases with the increase of the mean value of CFOs $E(\delta_f)$. While the dashed line in in Fig. \ref{fig:CFO_V} fluctuates a lot in high areas, which means that the performance of velocity estimation is poor when the variance of CFOs $V(\delta_f)$ is large.

\subsection{Sensing Performance Enhancement by Cooperative Sensing}
According to the analysis introduced in Section \ref{sec:Delay_Doppler_5}, after CCCS, cooperative sensing can improve the sensing performance by fusing active and passive sensing signals, which can be manifested in two aspects: target estimation and detection.

\subsubsection{NMSE of range and velocity of cooperative sensing after CCCS}\label{sec:CSCC_2}
The NMSE of range and velocity is an important metric for target estimation.
In this section, we compared the performance enhancement effect of cooperative sensing on active and passive sensing for different power ratios between them.
Finally, the NMSE of range and velocity of passive sensing after CCCS is analyzed and simulated to verify the improvement of passive sensing performance by CCCS.
\begin{figure*}[ht]
	\centering
	\subfigure[\scriptsize{MSE of $D_R(n)$ versus power ratio $\rm SNR_r$}.]{\includegraphics[width=0.45\textwidth]{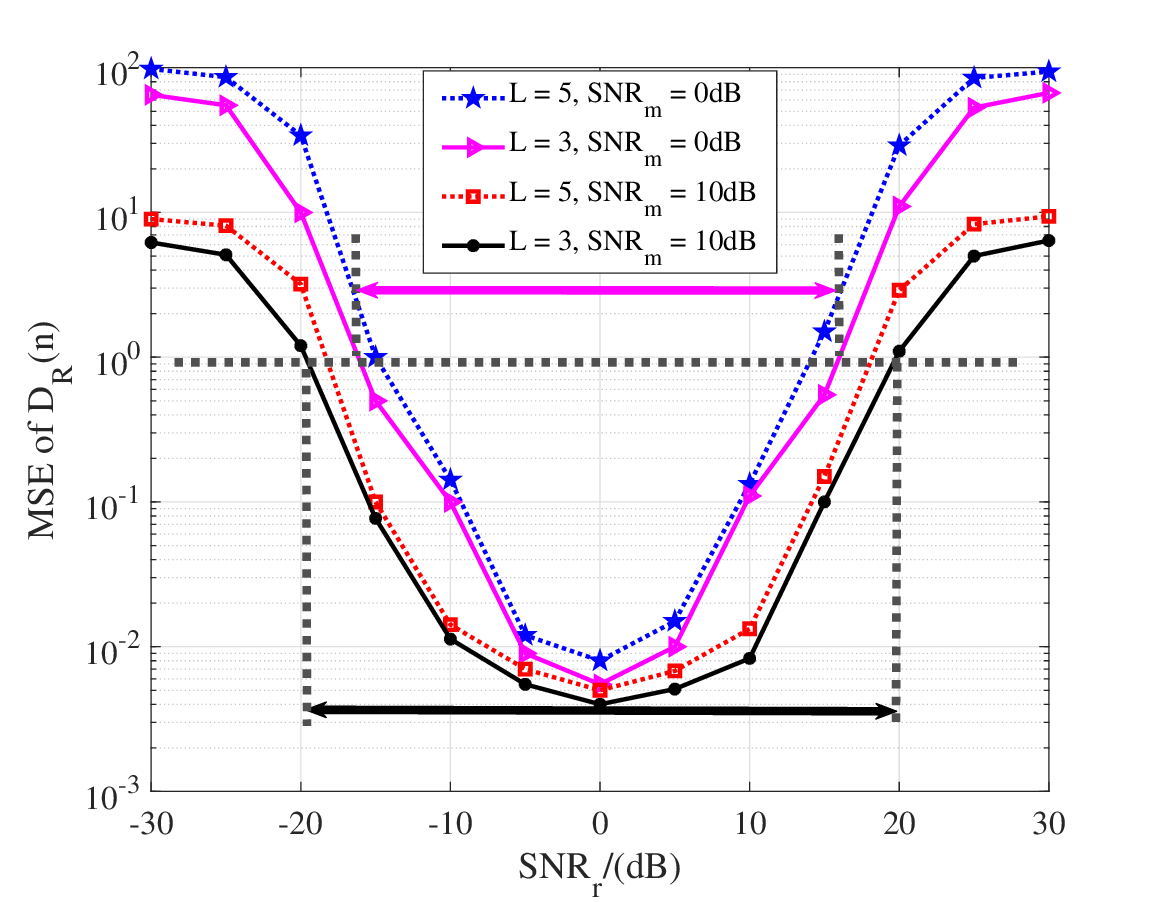}%
		\label{fig:MSE_DR}}
	\hfil
	\subfigure[\scriptsize{NMSE of range with $\rm SNR_m = 0 dB$, $L = 3$ }.]{\includegraphics[width=0.45\textwidth]{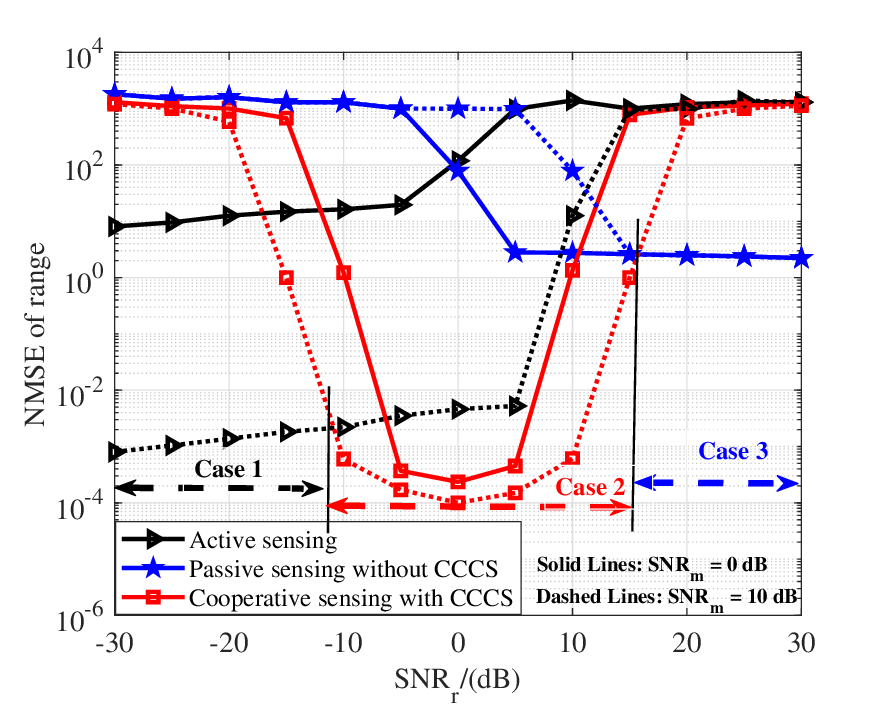}%
		\label{fig:NMSE_R_C}}
	\hfil
	\caption{Sensing performance enhancement of cooperative sensing for different power ratios $\rm SNR_r$ with $E(\delta_f)$ = 0.2 $\Delta f$, $E(\delta_\tau)$ = 100 ns, $V(\delta_f)$ = 0.01 and $V(\delta_\tau)$ = 100. }
	\label{fig:NMSE_R_C_DR}
\end{figure*}
MSE of $D_{R}(n)$ and $D_{V}(m)$ mentioned in Section \ref{sec:Delay_Doppler_4} reflects the accuracy of the CCCS outputs and directly impacts the sensing performance including NMSE of range $\Gamma_{R}$ and NMSE of velocity $\Gamma_{V}$.
According to Section \ref{sec:Delay_Doppler_4}, $D_{R}(n)$ is obtained by correlating the active sensing and passive sensing, which means that the power ratio between the two types of sensing, defined as $\rm SNR_r = \frac{\rm SNR_b}{\rm SNR_m}$, may affect the accuracy of $D_{R}(n)$.
Fig. \ref{fig:MSE_DR} presents MSE of $D_{R}(n)$ versus $\rm SNR_r$ with different $\rm SNR_m$ and number of targets.
We can see that MSE of $D_{R}(n)$ decreases and then increases as $\rm SNR_r$ increases. MSE of $D_{R}(n)$ is relatively smaller when the absolute value of $\rm SNR_r$ is smaller, i.e., the power of active sensing is closer to that of passive sensing, which implies a higher estimation accuracy.
By comparing the curves under different $\rm SNR_m$ in Fig. \ref{fig:MSE_DR}, it can be found that when the $\rm SNR_m$ is larger, the wider the range of $\rm SNR_r$ that satisfies the MSE of $D_{R}(n)$ less than a defined threshold, such as $1$, which means that the larger the $\rm SNR_m$, the CCCS can achieve TO and CFO estimation at a larger $|\rm SNR_r|$, and the system is more robust.

We will further analyze the performance enhancement effect of the cooperation based joint active and passive sensing for different $\rm SNR_r$.
Without loss of generality, assuming that $\rm SNR_m$ = 0 dB, $L$ = 3, $E(\delta_f)$ = 0.2 $\Delta f$, $E(\delta_\tau)$ = 100 ns, $V(\delta_f)$ = 0.01 and $V(\delta_\tau)$ = 100.
We take NMSE of range as an example to analyze the improvement of cooperative performance under different power ratios $\rm SNR_r$.
As Fig. \ref{fig:NMSE_R_C} shows, we can see that:
\begin{itemize}
\item For active sensing, where the received signal of passive sensing is considered interference.
When $\rm SNR_r$ is small, NMSE of range $\Gamma_{R}$ is relatively low. With $\rm SNR_r$ increases, gradual enhancement of passive sensing received signals will interfere with active sensing received signals and causes $\Gamma_{R}$ to be larger.
\item For passive sensing, where the received signal of active sensing is considered interference.
When $\rm SNR_r$ is small, $\Gamma_{R}$ is relatively high. With $\rm SNR_r$ increases, gradual enhancement of passive sensing received signals makes $\Gamma_{R}$ become smaller. However, $\Gamma_{R}$ for passive sensing does not drop very low due to the impact of TOs and CFOs.
\item For cooperation based joint active and passive sensing, where two sensing signals are fused by CCCS.
When $\rm SNR_r$ is small, i.e., $\rm SNR_b$ is small, $\Gamma_{R}$ for cooperative sensing is relatively high, close to that for passive sensing. With $\rm SNR_r$ increases, $\Gamma_{R}$ for cooperative sensing is decreasing, lower than that for active sensing. But as $\rm SNR_r$ increases further, MSE of $D_{R}(n)$ increases as well, which has been analyzed above, leading to higher $\Gamma_{R}$ for cooperative sensing.
\end{itemize}
{\color {black}
Therefore, we can conclude that:
\begin{itemize}
	\item when the received signal power of active sensing is much higher than that of passive sensing, i.e., case 1 as shown in Fig. \ref{fig:NMSE_R_C}, the sensing performance of active sensing is the best.
	\item when the received signal power of active sensing is close to that of passive sensing, i.e., case 2 as shown in Fig. \ref{fig:NMSE_R_C}, the sensing performance of cooperation based joint active and passive sensing is the best.
	\item when the received signal power of active sensing is much lower than that of passive sensing, i.e., case 3 as shown in Fig. \ref{fig:NMSE_R_C}, the sensing performance of passive sensing is the best.
\end{itemize}
}
To further analyze the improvement of cooperative sensing after CCCS in NMSE of range and velocity, we simulate the NMSE of range and velocity with different TOs, CFOs and number of targets. Without loss of generality, assuming that $\rm SNR_r =$ 0 dB, $E(\delta_f)$ = 0.2 $\Delta f$, $E(\delta_\tau)$ = 100 ns and $V(\delta_\tau)$ = 100.
\begin{figure*}[ht]
	\centering
	\subfigure[\scriptsize{NMSE of range versus $\rm SNR_b$}.]{\includegraphics[width=0.45\textwidth]{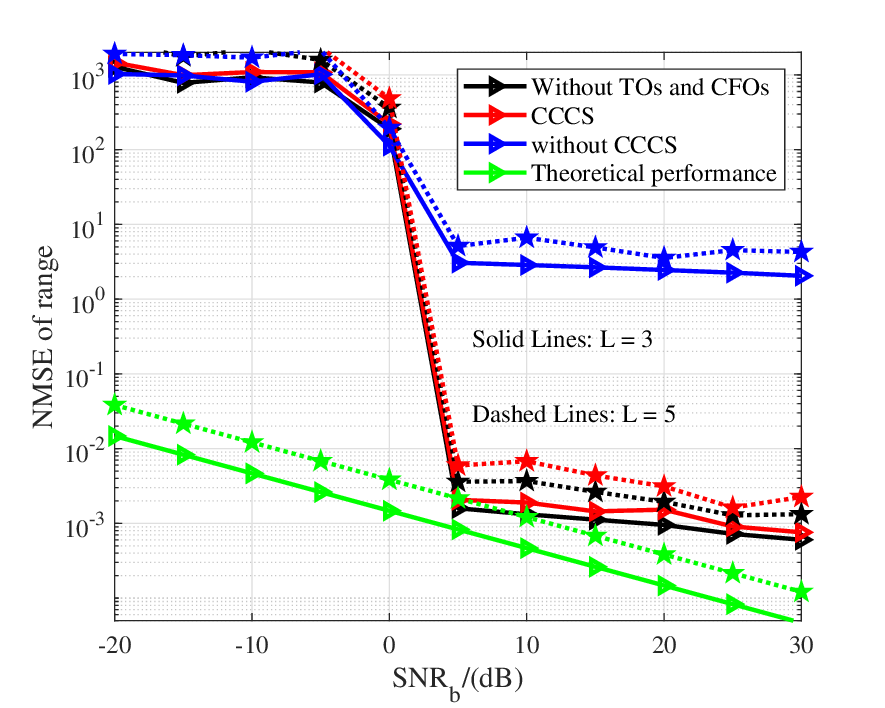}%
		\label{fig:NMSE_R}}
	\hfil
	\subfigure[\scriptsize{NMSE of velocity versus $\rm SNR_b$ }.]{\includegraphics[width=0.45\textwidth]{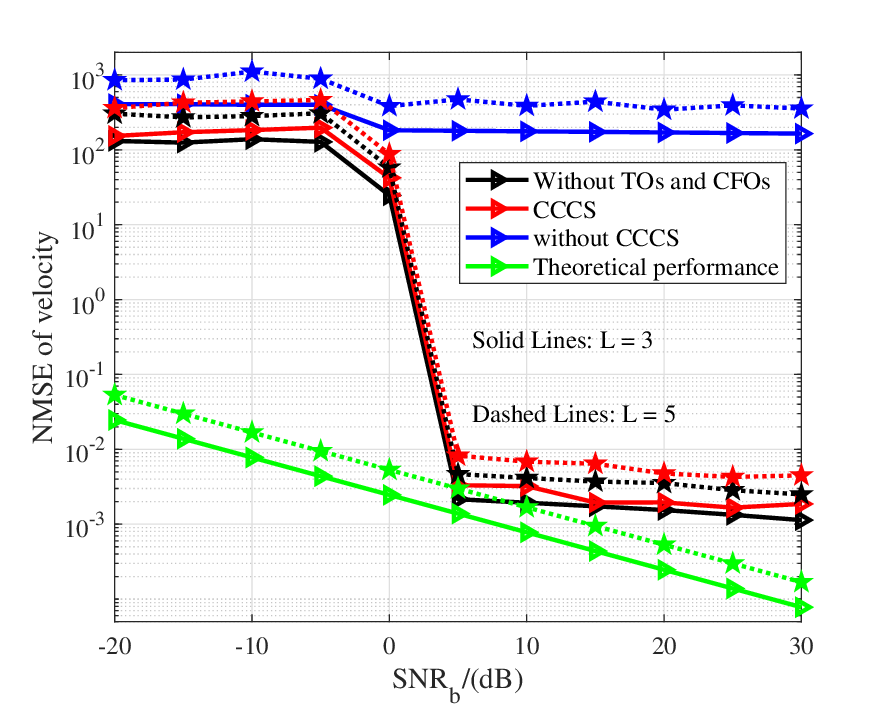}%
		\label{fig:NMSE_V}}
	\hfil
	\caption{NMSE of range and velocity versus $\rm SNR_b$ with $E(\delta_\tau)$ = 100 ns, $V(\delta_\tau)$ = 100, $E(\delta_f)$ = 0.2 $\Delta f$ and $V(\delta_f)$ = 0.01.}
	\label{fig:NMSE_R_V}
\end{figure*}
Fig. \ref{fig:NMSE_R_V} presents the NMSE of range and velocity versus $\rm SNR_b$, and we can get the following conclusions:
{\color {black}
\begin{itemize}
\item We can see that the dashed line of the same color is higher than the solid line, which indicates that an increase in the number of targets leads to an increase in NMSE of range and velocity and a consequent decrease in estimation accuracy.

\item We can see that the red line is well below the blue line and close to the black line, which means that CCCS can improve the passive sensing performance of range and velocity estimation under TOs and CFOs, close to the performance under perfect case, i.e., without TOs and CFOs.

\item The green line in Fig. \ref{fig:NMSE_R} denotes the theoretical performance of range estimation, which can be deduced by the standard deviation of Cramér–Rao bound, $\frac{\Delta R}{\sqrt{(2 \rm SNR_b)}}$ \cite{[TP]}, with $\Delta R = \frac{c}{2B}$ being the ranging resolution.
The green line in Fig. \ref{fig:NMSE_V} denotes the theoretical performance of velocity estimation, which can be deduced by the standard deviation of Cramér–Rao bound, $\frac{\Delta V}{\sqrt{(2 \rm SNR_b)}}$, with $\Delta V = \frac{c}{2TMf_{c2}}$ being the velocity resolution.
We can see that the red and black solid line is close to the green solid line when $\rm SNR_b = 5 dB$, the sensing performance of range and velocity estimation by CCCS is close to the theoretical performance when $\rm SNR_b = 5 dB$.

\end{itemize}
}
\begin{figure*}[ht]
	\centering
	\subfigure[\scriptsize{NMSE of velocity versus $V(\delta_f)$ with $\rm SNR_b$ = 10 dB, $E(\delta_f)$ = 0.2 $\Delta f$, $E(\delta_\tau)$ = 100 ns and $V(\delta_\tau)$ = 100}.]{\includegraphics[width=0.45\textwidth]{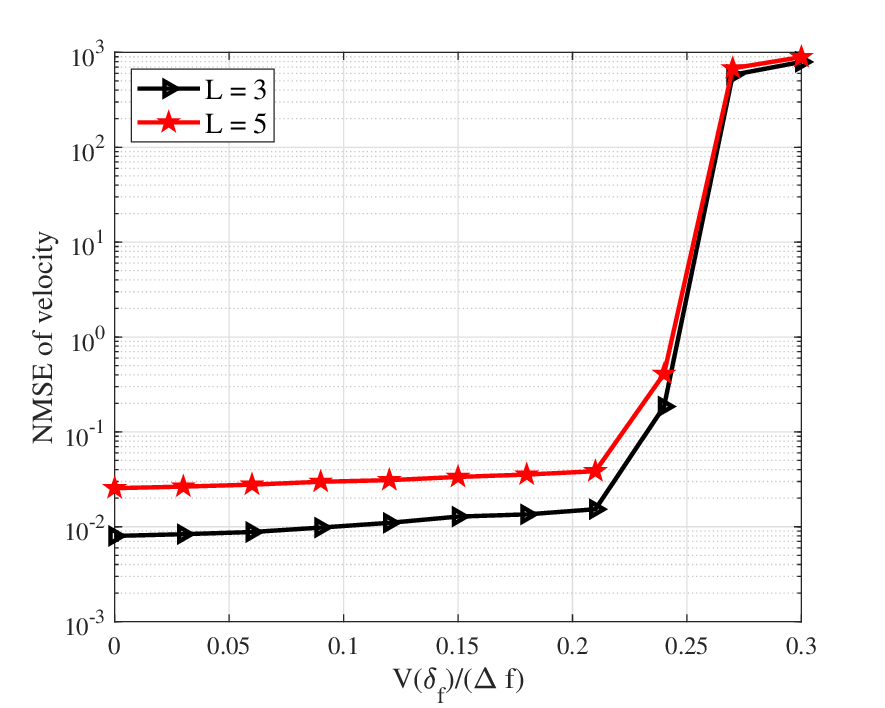}%
		\label{fig:NMSE_V_V}}
	\hfil
	\subfigure[\scriptsize{ROC for different type sensing with $\rm SNR_r$ = 0 dB }.]{\includegraphics[width=0.45\textwidth]{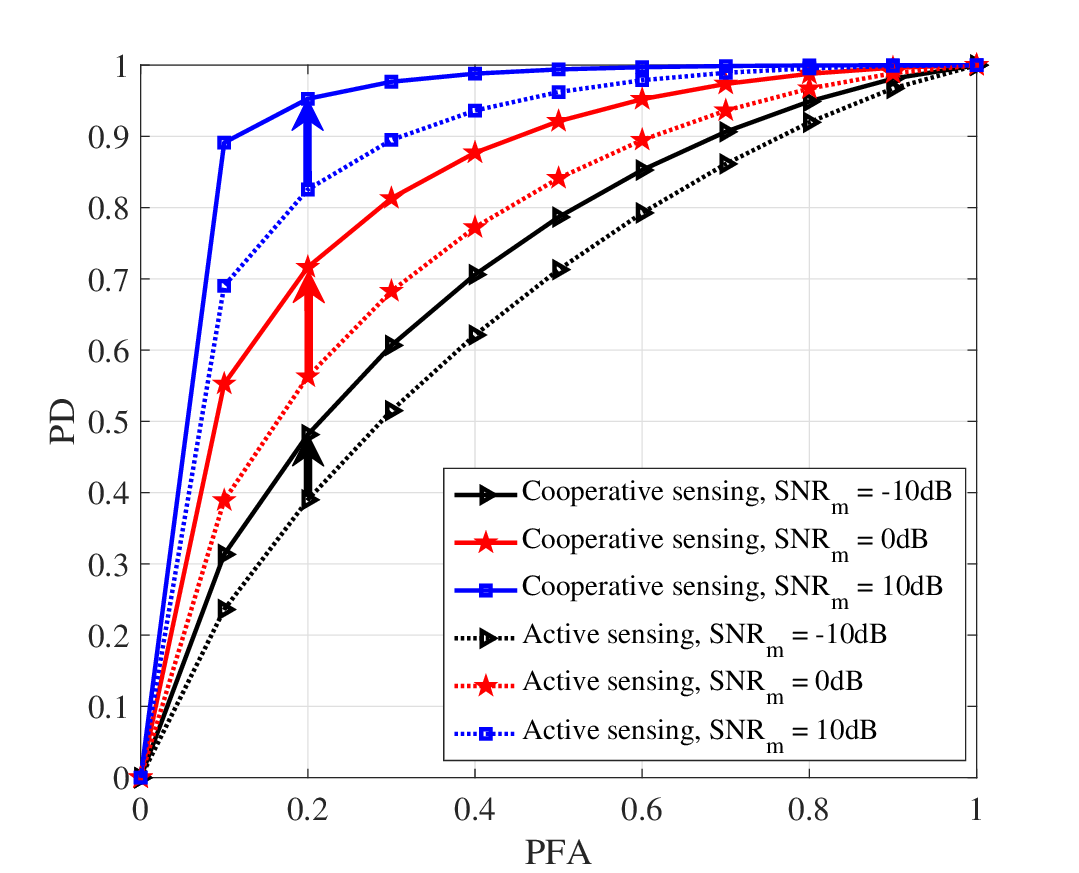}%
		\label{fig:ROC}}
	\hfil
	\caption{NMSE of velocity and ROC}
	\label{fig:6}
\end{figure*}

According to Fig. \ref{fig:CFO_V} and the analysis mentioned in Section \ref{sec:Delay_Doppler_4_2}, the variance of CFOs $V(\delta_f)$ results in that the Doppler spread on different OFDM symbols $m$ is variable, which in turn affects the accuracy of DFT for estimating $\Delta f_{\textrm{D},l} + \delta_f(m)$, further affects the performance of CCCS for mitigating CFOs, ultimately resulting in lower velocity estimation accuracy.
Recognizing this fact, we simulate the NMSE of velocity versus $V(\delta_f)$, as shown in Fig. \ref{fig:NMSE_V_V}. We can see that the performance of CCCS for mitigating CFOs is relatively great when $V(\delta_f)$ is smaller than 0.25, which can be obtained by ordinary frequency sources \cite{[CFO_S_3],[CFO_S_4],[CFO_S_5]}.

\subsubsection{ROC of cooperative sensing}\label{sec:CSCC_3}
In terms of target detection, the receiver operating characteristic (ROC) is an important metric, which denotes the relationship between the probability of detection (PD) and the probability of false alarm (PFA). Without loss of generality, assuming that $\rm SNR_r$ = 0 dB and Neyman-Pearson test principle is adopted \cite{[NP_test]}. As shown in Fig. \ref{fig:ROC}, the ROC of cooperative sensing is higher than that of active sensing under the same SNR, which indicates that cooperative sensing can improve the target detection performance of active sensing to a certain extent.

\subsection{AoA Estimation}\label{sec:Simulation_2}
{\color{black}
%%一个图，展示DFT-MUSIC估计的粗到精的示意图，
%Fig. \ref{fig:AOA_1} illustrates an example of the low complexity AOA estimation. Assuming that there are three targets with AOA of $25°$, $30°$ and $35°$. According to Fig. \ref{fig:AOA_1} and  \eqref{equ:AOA_1_1_3}, CAE results can be obtained by DFT
%\begin{equation} \label{equ:Simulation_2_1}
%\begin{aligned}
%\theta_{l,c} &\in [ \rm{acos}(\frac{\lambda P_i}{d N_t}), \rm{acos}( \frac{\lambda P_i}{d N_t} + \frac{\lambda}{2 \pi d}        )    ) *180 / \pi
%&\in ( 21.0575^{o}, 41.4096^{o} ]
%\end{aligned},
%\end{equation}
%where $N_t = 8$, $d = \lambda / 2$, $P_i$ is the index of the peak of DFT outputs, which is $3$ as shown in the below part of Fig. \ref{fig:AOA_1}.
%Then, FAE results can be obtained by using MUSIC to search within CAE, i.e., $( 21.0575^{o}, 41.4096^{o} ]$, just as shown in the upper part of Fig. \ref{fig:AOA_1}.
%\begin{figure}[ht]
%	\includegraphics[scale=0.45]{MSBS/AOA_2.eps}
%	\centering
%	\caption{An example of the low complexity AOA estimation.}
%	\label{fig:AOA_1}
%\end{figure}

%一个图，MMSE_AOA (本文算法，conventional MUSIC，ESPRIT, UL_sensing)
\begin{figure*}[ht]
	\centering
	\subfigure[\scriptsize{RMSE of AoA with variable SNR}.]{\includegraphics[width=0.45\textwidth]{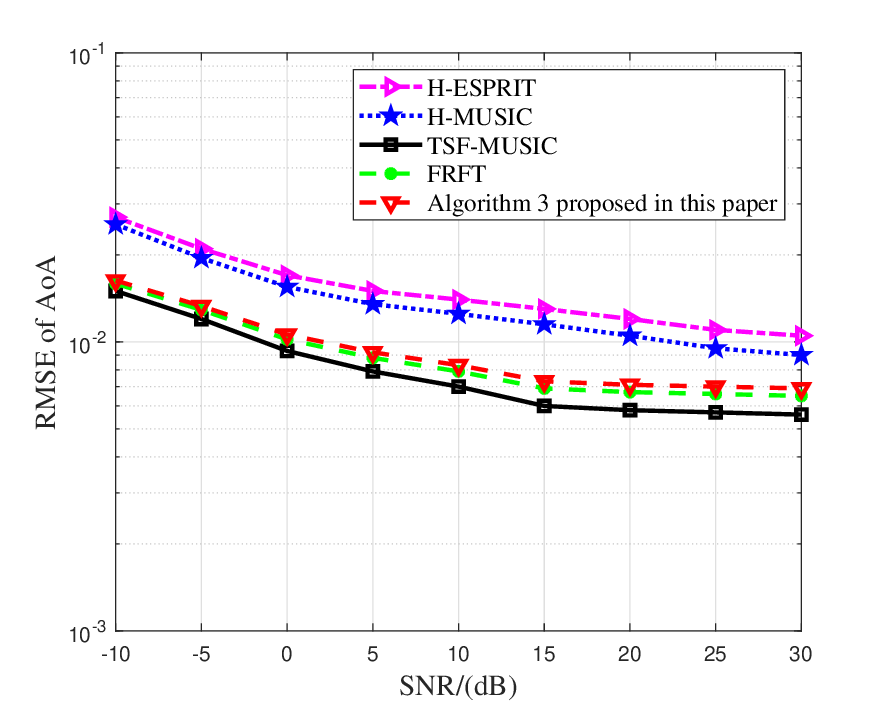}%
		\label{fig:RMSE_AOA}}
	\hfil
	\subfigure[\scriptsize{Complexity of different AoA estimation method }.]{\includegraphics[width=0.45\textwidth]{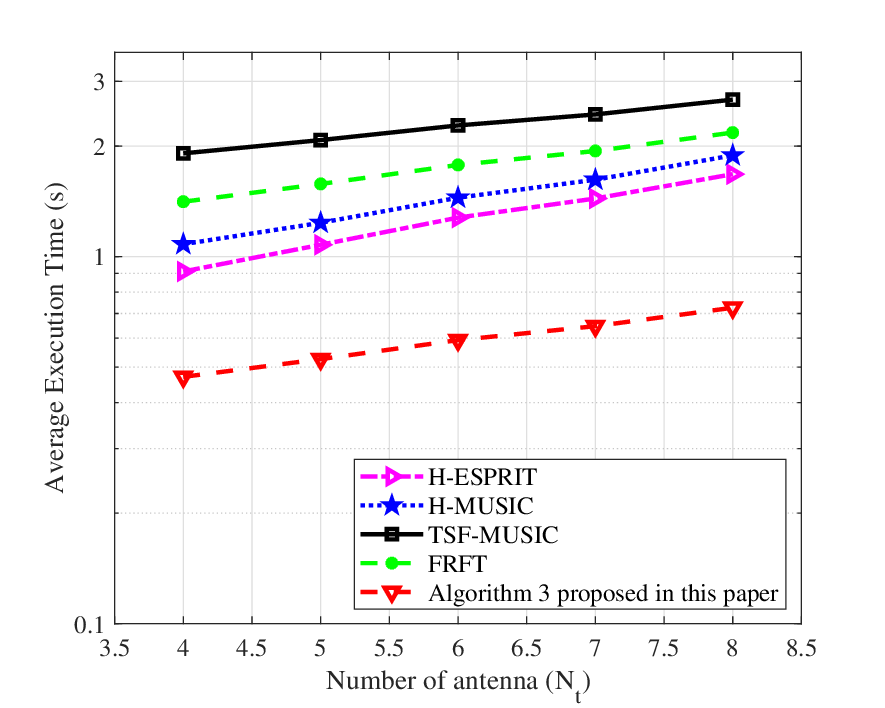}%
		\label{fig:Com}}
	\hfil
	\caption{RMSE of AoA with variable SNR and Complexity of different AoA estimation method. }
	\label{fig:AOA}
\end{figure*}
Fig. \ref{fig:RMSE_AOA} shows the root mean square error (RMSE) of the AoA estimation with variable SNR, which can be expressed as
\begin{equation} \label{equ:Simulation_2_2}
\begin{aligned}
\Gamma_\textrm{A} &= \sqrt {\frac{1}{L} \sum_{l=0}^{L-1} {\left|\Omega_{e,l} - \Omega_{r,l}\right|^2}}
\end{aligned},
\end{equation}
where $\Omega_{r,l}$ and $\Omega_{e,l}$ are the actual and estimated AoA of targets, respectively.
We compare the proposed low complexity AoA estimation algorithm, i.e., Algorithm \ref{alg:FRFT}, with the FRFT from \cite{[QHY]}, the H-ESPRIT and H-MUSIC method from \cite{[AOA_29]}, and the TSF-MUSIC method from \cite{[UL_sensing]}.
\begin{itemize}
	\item The fractional index of FRFT is set as $N_f = 10$, the uncertainty range of CAE is set as $\Omega = \frac{\pi}{3}$.
	\item The setup of MUSIC is the same as that in TSF-MUSIC method \cite{[UL_sensing]}. %, $C = 128$ and $C_1 = 10$
\end{itemize}
%There are $3$ targets with the AoA of $25°$, $30°$ and $35°$, and
According to Fig. \ref{fig:RMSE_AOA}, RMSE of AoA, $\Gamma_\textrm{A}$ is close to that of FRFT, which is lower than that of H-ESPRIT and H-MUSIC \cite{[AOA_29]}, but little higher than that of TSF-MUSIC \cite{[UL_sensing]}, which means that the AoA estimation performance of Algorithm \ref{alg:FRFT} is close to FRFT \cite{[QHY]}.
In Fig. \ref{fig:Com}, we explore the complexity of different methods by the average execution time, which reflects the average time to complete AoA estimation for different methods. We can see that the red line is much lower than other lines, which means that the complexity of Algorithm \ref{alg:FRFT} is much lower than other algorithms.
Therefore, simulation results show that Algorithm \ref{alg:FRFT} can achieve relatively high-performance AoA estimation, close to FRFT, at a low complexity, much lower than FRFT.
}

\section{Conclusion}\label{sec:Conclusion}

In this paper, we propose a cooperation based joint active and passive sensing scheme to improve the sensing performance in PMN.
To overcome the clock asynchronicity problem in passive sensing, which is caused by the spatially separated asynchronous transceivers, we propose the CCCS based algorithm that mitigates TOs and CFOs by correlating active and passive sensing information.
Compared with most existing algorithms, the CCCS based algorithm is more widely applicable because it does not require the existence of LOS propagation paths between transceivers.
Moreover, we propose a low complexity algorithm that adopts coarse and fine precision iterative estimation to achieve high-accuracy AoA estimation with low complexity.
Finally, we analyze and simulate the performance improvement of the cooperation based joint active and passive sensing in delay and Doppler spread estimation.
The simulation results verify the CCCS based algorithm and the performance improvement of cooperative sensing.

\begin{appendices}

%\section{CCCS for Single Target} \label{app:A}
%
%For mitigating $\Delta \tau + \delta_\tau(m)$, the CCCS algorithm between ${\bf k}_{1,R}$ and ${\bf k}_{2,R}$ generates
%\begin{equation} \label{equ:A_1}
%\begin{aligned}
%{\bf \rho}_{R} &=   {\rm{diag}} ({\bf k}_{2,R}^{\textrm H}) \cdot {\bf k}_{1,R}
%\end{aligned},
%\end{equation}
%\begin{equation} \label{equ:A_2}
%\begin{aligned}
%{\bf k}_{1,R} &=
%{\begin{bmatrix}
%1,& \cdots,& e^{-j2 \pi n \Delta f \tau_{1}},& \cdots,&   e^{-j2 \pi (N-1) \Delta f \tau_{1}}
%\end{bmatrix}}^\textrm{T}
%,\in {{\mathcal C}^{N \times 1}}
%\end{aligned}
%\end{equation}
%\begin{equation} \label{equ:A_3}
%\begin{aligned}
%{\rm{diag}} ({\bf k}_{2,R}^{\textrm H}) &=
%\begin{bmatrix}
%1 &  & \\
%& \ddots & \\
%&  & e^{j2 \pi (N-1) \Delta f (\tau_{2} + \delta_\tau(m))}
%\end{bmatrix},,\in {{\mathcal C}^{N \times N}}
%\end{aligned}.
%\end{equation}
%Then ${\bf \rho}_{R}$ can be derived as
%\begin{equation} \label{equ:A_4}
%\begin{aligned}
%{\bf \rho}_{R} &=
%{\begin{bmatrix}
%1,& \cdots,& e^{j2 \pi n \Delta f (\tau_2 - \tau_1 + \delta_\tau(m))},& \cdots,&   e^{j2 \pi (N-1) \Delta f (\tau_2 - \tau_1 + \delta_\tau(m))}
%	\end{bmatrix}}^\textrm{T}
%\\ &=
%{\begin{bmatrix}
%	1,& \cdots,& e^{j2 \pi n \Delta f (\Delta \tau + \delta_\tau(m))},& \cdots,&   e^{j2 \pi (N-1) \Delta f (\Delta \tau + \delta_\tau(m))}
%	\end{bmatrix}}^\textrm{T}
%,\in {{\mathcal C}^{N \times 1}}
%\end{aligned}.
%\end{equation}
%
%In the similar way, ${\bf \rho}_{D}$, ${\bf k}_{2,R}^{'}$ and ${\bf k}_{2,D}^{'}$ can be derived.

\section{CCCS for Multiple Targets} \label{app:B}
According to \eqref{equ:CSCC_4_2_3} and \eqref{equ:CSCC_4_2_5}, the promotion from $D_{R}(n,k)$ to $D_{R}(n)$ can be derived as
\begin{equation} \label{equ:B_1}
\begin{aligned}
w_{D} &= \frac{D_{R}(n)}{D_{R}(n,k)} = N_t
\end{aligned}.
\end{equation}
According to \eqref{equ:CSCC_4_2_4} and \eqref{equ:CSCC_4_2_6}, the promotion from $I_{R}(n,k)$ to $I_{R}(n)$ can be derived as
\begin{equation} \label{equ:B_2}
\begin{aligned}
w_{I} &= \frac{I_{R}(n)}{I_{R}(n,k)} = \vec \sum_{k=0}^{N_t-1} {e^{j k \left(\Omega_{l_2} - \Omega_{l_1}\right) }}
\end{aligned},
\end{equation}
where $l_1 \ne l_2$, then $\Omega_{l_2}$ is not equal to $\Omega_{l_1}$ in most cases.
Then,
\begin{equation} \label{equ:B_3}
\begin{aligned}
 w_{I} = \vec \sum_{k=0}^{N_t-1} {e^{j k \left(\Omega_{l_2} - \Omega_{l_1}\right) }} < N_t
\end{aligned}.
\end{equation}
Therefore, $D_{R}(n,k)$ can be enhanced by calculate the vector sum of $\beta_{R}(n,k)$ on the $N_t$ antennas.

\end{appendices}

% if have a single appendix:
%\appendix[Proof of the Zonklar Equations]
% or
%\appendix  % for no appendix heading
% do not use \section anymore after \appendix, only \section*
% is possibly needed

% use appendices with more than one appendix
% then use \section to start each appendix
% you must declare a \section before using any
% \subsection or using \label (\appendices by itself
% starts a section numbered zero.)
%

% reference
\bibliographystyle{IEEEtran}
\bibliography{reference}

% use section* for acknowledgment

% Can use something like this to put references on a page
% by themselves when using endfloat and the captionsoff option.
\ifCLASSOPTIONcaptionsoff
  \newpage
\fi

\end{document}